\documentclass[12pt]{article}

\usepackage{amsfonts}

\newcommand{\rmi}{{\rm i}}
\newcommand{\rme}{{\rm e}}
\newcommand{\eref}[1]{(\ref{#1})}
\newcommand{\sref}[1]{section~\ref{#1}}
\newcommand{\Sref}[1]{Section~\ref{#1}}
\newcommand{\tref}[1]{table~\ref{#1}}

\newcommand{\br}{\ms\hline\ms}

\newcommand{\mr}{\ms\hline\ms}
\newcommand{\ms}{\noalign{\vspace{3pt plus2pt minus1pt}}}
\newcommand{\bs}{\noalign{\vspace{6pt plus2pt minus2pt}}}

\newcommand{\nonum}{\item[]}

\begin{document}

\title{
Two Dimensional Conformal Field Theory on Open and Unoriented Surfaces 
\footnote{Lectures presented the 4th SIGRAV 
Graduate School and 2001 School on Algebraic 
Geometry and Physics  ``Geometry and Physics of Branes'' in Como, Italy, May 2001.}
}
 
\author{ 
\Large{Yassen  S. Stanev} 
\\
\\
{\it Dipartimento di Fisica, \ Universit{\`a} di Roma \  ``Tor Vergata''} \\  
{\it I.N.F.N.\ -\ Sezione di Roma \ ``Tor Vergata''} \\ 
{\it Via della Ricerca  Scientifica 1, 00133 \ Roma, \ Italy} \\ 
{\it and } \\
{\it Institute for Nuclear Research and Nuclear Energy} \\
{\it Bulgarian Academy of Sciences, BG-1784 \ Sofia, \ Bulgaria} 
}

\maketitle

\begin{abstract}
Introduction to two dimensional conformal field theory on 
open and unoriented surfaces. The construction is illustrated 
in detail on the example of SU(2) WZW models.
\end{abstract}

\newpage

\section{Introduction}

Two dimensional Conformal Field Theory (CFT) on open and unoriented surfaces 
is not a recent discovery. Its systematic study began in two seemingly different 
developments.
On the one hand the implications of the 
presence of a boundary in two dimensional systems 
and the corresponding boundary conditions and boundary 
fields were first analyzed  by  Cardy \cite{[93]} and further in \cite{[94],[97]}.
On the other hand a general  prescription for the systematic construction of open and unoriented string models 
from a given closed oriented string model was proposed by Sagnotti \cite{[92]} and 
further elaborated in \cite{[103],[104]}. However it was only after 
the discovery of D-branes \cite{[PolchD]} that the topic attracted so much attention
and a huge number of different  models have been explicitly constructed (any list will
be incomplete). 
A parallel development was the study of the general consistency conditions for the
models, and in particular of the compatibility conditions between the Klein bottle 
projection and the annulus partition function embodied by the M\"obius 
strip projection. As often in two dimensional conformal theories a rational 
completely solved model like the $SU(2)$ Wess--Zumino--Witten  model
provided a good playground for such an analysis and exhibited 
three interesting properties

- for the diagonal models there is a standard solution 
which extends the Cardy ansatz for the annulus
to the unoriented case \cite{[101]};

- there may be  several different  Klein bottle projections 
corresponding to different spectra in the unoriented sector;

- the annulus partition function satisfies a completeness condition 
(satisfies the chiral fusion algebra) \cite{[98]}.

The last property extends also to all other explicitly solved examples, but
a better understanding of the physical principle underlying the 
completeness condition in the general case, 
in particular in the framework of
string theory where so far open and closed string completeness conditions appear rather asymmetrically, 
is still absent. 
Another important open problem is whether there will be new constraints on the unoriented sector
coming from higher genus surfaces.

Two dimensional conformal field theory on surfaces with boundaries
and crosscaps is a large and rapidly developing subject. 
The aim of these lectures is to give an  introduction to the topic,
hence we have chosen to present a self-contained exposition based on  
one relatively simple and completely solved example, namely the  
$SU(2)$ Wess--Zumino--Witten  (WZW) model.
Even so some  aspects like the 
explicit realization of the models in terms of D-branes and orientifolds
\cite{[D-branes]} and their geometry are not covered.
Other important developments which have to be mentioned are the
relations of boundary conformal theory to graph theory (for a review see 
\cite{[ZuberReview]}) and to topological field theory \cite{[Topological]}.

The material is organized as follows.
In \sref{sec2} we review some general properties of two dimensional CFT.
In  \sref{sec3} we derive explicit expressions for the 4-point functions in the 
$SU(2)$ WZW model, the corresponding exchange operators and fusion matrix. 
 \Sref{sec4} is devoted to the derivation of the sewing constraints for the
correlation functions on open and unoriented surfaces.
In  \sref{sec5} we analyze the partition functions and the consistency 
conditions they satisfy.

\section{General properties of two dimensional CFT}
\label{sec2}

\subsection{The stress energy tensor in two dimensions}  

Let us begin by recalling   the particular properties of the 
stress energy tensor in  two dimensional conformal field theory. 
It is useful to introduce together
with the flat Minkowski space light cone coordinates $x_{\pm}= x^0 \pm x^1$ 
also the coordinates on the cylindric space ${\mathbb S}^1 \times {\mathbb R}^1$
(on which the conformal transformations are well defined globally \cite{[2]})
$t_\pm = \xi^0 \pm \xi^1$. Here $\xi^0$ is the non-compact time variable
on the cylinder, while $\xi^1$ is the compact space variable 
($\xi^1+2\pi$ is identified with $\xi^1$).
We shall use also the  analytic  picture on the compact space  
${\mathbb S}^1 \times {\mathbb S}^1$ with coordinates 
\begin{equation}
z= \rme^{\rmi t_-} \  \qquad {\bar z}= \rme^{\rmi t_+} \quad , 
\label{eq:1.14}
\end{equation}
where the complex variables $z$ and $\bar z$ are obtained from the Minkowski 
light cone coordinates by a Cayley transform 
\begin{equation}
z={1+{\rmi \over 2} x_- \over 1-{\rmi \over 2}x_-} \  \qquad 
{\bar z} ={1+{\rmi \over 2} x_+ \over 1-{\rmi \over 2}x_+} \quad .
\label{eq:1.15}
\end{equation}
Note that $z$ and $\bar z$ are complex conjugate 
only if one starts from the Euclidean picture where $\xi^0$ 
is purely imaginary, while $\xi^1$ is real.
Nonlinear transformations of the coordinates, like \eref{eq:1.15},
require nontrivial accompanying changes of the field variables.
To find the transformation law for the stress energy tensor let us first write 
its components in the light cone basis $x_\pm$
\begin{eqnarray}
\Theta_{\mu \nu} dx^\mu dx^\nu &=& 
\Theta_{++}dx_+^2 +\Theta_{+-}dx_+dx_- \nonumber \\
&+& \Theta_{-+}dx_-dx_+
+\Theta_{--}dx_-^2 \quad , 
\label{eq:1.16a}
\end{eqnarray}
where
\begin{eqnarray}
 \Theta_{++} & =&  {1 \over 4 }
\left( \Theta_{00} +\Theta_{10} +\Theta_{01} +\Theta_{11} \right) \   \nonumber \\
 \Theta_{--} & =&  {1 \over 4 }
\left( \Theta_{00} -\Theta_{10} -\Theta_{01} +\Theta_{11} \right) \  \nonumber \\
 \Theta_{+-} & =&   \Theta_{-+}  \ = \   {1 \over 4 }
\left( \Theta_{00} -\Theta_{11} \right)  \ . \nonumber 
\label{eq:1.16b}
\end{eqnarray}
The energy density with our choice of metric
\begin{equation}
 \eta_{\mu \nu} = {\rm diag( - , +) }
\label{eq:16d}
\end{equation}
 is given by
 $\Theta_0^0 = -\Theta_{00}$, so let us choose the three independent components
of $\Theta_{\mu \nu} $ as 
\begin{equation}
\Theta = -\Theta_{--} \  \qquad {\bar \Theta} = - \Theta_{++}  
\qquad \Theta_0 = - \Theta_{+-} = {1 \over 4} {\rm Tr} \Theta  \ .
\label{eq:1.17}
\end{equation}
The conservation of the stress energy tensor 
 $\partial_\mu \Theta^{\mu \nu} = 0 $ then implies
\begin{equation}
\partial_+ \Theta = - \partial_- \Theta_0 \ \qquad 
\partial_- {\bar \Theta} = - \partial_+ \Theta_0 \ ,
\label{eq:1.18}
\end{equation}
where $ \partial_\pm = 1/2(\partial_0 \pm \partial_1)$.
The corresponding fields in the analytic picture are 
\begin{eqnarray}
T(z,\bar z) & = & 2 \pi \left( \rmi{\partial x_- \over \partial z}\right)^2
\Theta(x_+(\bar z),x_-(z)) \  \nonumber \\
\ms
\bar T(z,\bar z) & = &  2 \pi \left( \rmi{\partial x_+ \over \partial \bar z}\right)^2
\bar \Theta(x_+(\bar z),x_-(z)) \  \\
\ms
T_0(z,\bar z) & = & 2 \pi \left( \rmi{\partial x_- \over \partial z}\right)
\left( \rmi{\partial x_+ \over \partial \bar z}\right)
\Theta_0(x_+(\bar z),x_-(z)) \ . \nonumber 
\label{eq:1.19}
\end{eqnarray}
The conservation of $\Theta$ leads to the equations 
\begin{equation}
\bar \partial T = - \partial T_0 \  \qquad 
\partial \bar T = - \bar \partial T_0 \  \qquad 
\left( \partial = {\partial \over \partial z}, \bar \partial = {\partial \over  \partial \bar
z} \right) \ . 
\label{eq:1.20}
\end{equation}
Thus if the stress energy tensor is traceless
 $(T_0 = 0)$ each of the two components  $T$ and  $\bar T$
depends on a single variable  $T = T(z)$ and 
$\bar T = \bar T(\bar z)$. 

A similar  separation in chiral and antichiral components
is valid also for an abelian current
$j_\mu$  that is conserved together with its dual
\begin{equation}
\partial_\mu j^\mu = 0 = \partial^\mu \epsilon_{\mu \nu} j^\nu .
\label{eq:1.21}
\end{equation}

We shall call such fields which split into chiral and antichiral 
components local observables. In other words, one can define the 
two dimensional conformal field theory as a quantum field theory 
in which the observable algebra is a tensor product of two algebras 
\begin{equation}
{\cal A} \otimes \bar {\cal A}  \ .
\label{eq:1.22}
\end{equation}
The chiral (or analytic) algebra ${\cal A}$
\index{algebra ! chiral} and the 
antichiral (or antianalytic) algebra $\bar {\cal A}$
are related by space reflection.
For the rest of these lectures we shall assume 
that ${\cal A}$ and $\bar {\cal A}$  are isomorphic. 
The algebra  $\cal A$ is generated by a finite number of local fields 
$O_n(z)$. It should be stressed that this condition does not lead necessarily to 
a finite number of fields in the theory.
Locality implies that all $O_n(z)$ mutually commute for 
different arguments, more precisely for any given $n$ and $m$ 
there exists an integer $N_0(n,m)$ such that for all
$ N \geq N_0$
\begin{equation}
(z_1 - z_2)^N \left[ O_n(z_1), O_m(z_2) \right] = 0 \ .
\label{eq:1.23}
\end{equation}
The general solution of this equation is given by a linear combination of 
the $\delta$ function and its derivatives 
\begin{equation}
\left[ O_n(z_1), O_m(z_2) \right] = \sum_{\ell=0}^{N_0-1} C_{\ell}(z_2) 
\delta^{(\ell)}(z_{12}) \ ,
\label{eq:1.24}
\end{equation}
where $\delta$ on the unit circle can be defined as  
\begin{eqnarray}
\delta(z_{12})  & = &  {1 \over z_1} \sum_n \left( {z_2 \over z_1} \right)^n
 \nonumber \\
& = &{1 \over z_1} \sum_{n=0}^{\infty} \left( {z_2 \over z_1} \right)^n +
 {1 \over z_2} \sum_{n=0}^{\infty} \left( {z_1 \over z_2} \right)^n \quad  
\label{eq:1.25a}
\end{eqnarray}
and satisfies
\begin{equation}
\oint \delta(z_{12}) f(z_2) {dz_2 \over 2 \pi i} = f(z_1) \ .
\label{eq:1.25b}
\end{equation}

For the currents (of scale dimension 1) and for the stress energy 
tensor (of scale dimension 2) this leaves undetermined only one constant.
In particular \cite{[10]}
\begin{equation}
\left[T(z_1),T(z_2)\right]   =   -{c \over 12} \delta^{'''}(z_{12}) -
\delta^{'}(z_{12})(T(z_1) + T(z_2))  \ , 
\label{eq:1.26}
\end{equation}
where the constant $c$ is called central charge.
The same relation holds also for the antichiral component 
$\bar T$ with central charge $\bar c$ ($=c$ due to the assumption that 
${\cal A}$ and  $\bar {\cal A}$ are isomorphic).
All fields from ${\cal A}$ commute with 
all fields from $\bar {\cal A}$, hence  
$T(z)$ and $\bar T(\bar z)$ commute.
Under a general analytic reparametrization $z \rightarrow w(z)$
the stress energy tensor transforms according to  
\begin{equation}
T(z) \rightarrow T(w) = \left( {\partial z \over \partial w} \right)^2
T(z(w))  + {c \over 12} \left\{ w , z \right\} \ , 
\label{eq:1.27}
\end{equation}
where  $\{w,z\}$ is the Schwartz derivative
\begin{equation}
\{ w , z \} = {w^{'''} \over w^{'}} - 
{3 \over 2} \left( w^{''} \over w^{'} \right)^2 .
\label{eq:1.28}
\end{equation}
The central term in  \eref{eq:1.26},\eref{eq:1.27} is related to the conformal anomaly.
$T(z)$  has a  Laurent expansion of the form 
\begin{equation}
T(z) = \sum_n { L_n \over  z^{n+2} } \ ,
\label{eq:1.29}
\end{equation}
where the modes $L_n$ are given by  
\begin{equation}
L_n = {1 \over 2 \pi i } \oint_{S^1} dz \ T(z) \  z^{n+1} \quad .
\label{eq:1.30}
\end{equation}
The commutator \eref{eq:1.26} for the chiral components of the 
stress energy tensor implies for the modes $L_n$ 
the commutation relations of the Virasoro algebra \index{algebra ! Virasoro}
$Vir$ \cite{[9]} 
\begin{equation}
\left[ L_n , L_m \right] = (n-m) L_{n+m} +{ c \over 12} n(n^2-1) 
\delta_{n+m} \ ,  
\label{eq:1.31}
\end{equation}
where $\delta_{\ell}$ denotes the  Kronecker symbol $\delta_{\ell, 0}$.
The central term in \eref{eq:1.31} vanishes for $n=0,\pm 1$.
The corresponding subalgebra generated by $L_{-1}$, $L_0$ and $L_1$
is $SL(2,\mathbb R)$.
The unique vacuum vector $\vert 0 \rangle$ is annihilated by $L_{-1}$, $L_0$ and $L_1$
 (and by their antichiral counterparts) 
\begin{equation}
L_{0,\pm 1} \vert 0 \rangle = 0  = \bar L_{0,\pm 1} \vert 0 \rangle \ .
\label{eq:1.32}
\end{equation}
The Hermiticity of the stress energy tensor gives for the modes 
\begin{equation}
L_n^{\dagger} = L_{-n} \quad  . 
\label{eq:1.33}
\end{equation}

Not all the fields in the theory split into chiral and antichiral parts.
In particular there exist ``primary'' conformal fields \cite{[10],[11]}, 
of conformal weights $\Delta$ and $\bar \Delta$, which under reparametrizations
$z \rightarrow w(z)$, $\bar z \rightarrow \bar w(\bar z)$
transform as \index{field ! primary}
\begin{equation}
\phi_{\Delta \bar \Delta}(z, \bar z) \rightarrow \phi_{\Delta \bar \Delta}(w,
\bar w) = 
\left( {\partial z \over \partial w} \right)^{\Delta}
\left( {\partial \bar z \over \partial \bar w} \right)^{\bar \Delta}
\phi_{\Delta \bar \Delta}(z(w),\bar z (\bar w)) \ .
\label{eq:1.34}
\end{equation}
This transformation law implies the following commutation relations
between the primary fields and the generators of the Virasoro algebra
$L_n$
\begin{eqnarray}
\left[ L_n , \phi_{\Delta \bar \Delta}(z,\bar z) \right] & = & z^n \left( z
\partial_z +(n+1)\Delta
\right)  \phi_{\Delta \bar \Delta}(z,\bar z) \  
\label{eq:1.35a} \\
\ms
\left[ \bar L_n , \phi_{\Delta \bar \Delta}(z,\bar z) \right] & = & \bar z^n
\left( \bar z
\partial_{\bar z} +(n+1)
\bar \Delta
\right)  \phi_{\Delta \bar \Delta}(z,\bar z) \ . 
\label{eq:1.35b}
\end{eqnarray}
The corresponding states obtained by acting with the 
primary fields on the vacuum are also called primary
\index{state ! primary}
\begin{equation}
\vert \Delta , \bar \Delta \rangle \ = \ \phi_{\Delta \bar \Delta}(0,0) \ 
\vert 0  \rangle \ .
\label{eq:1.35c}
\end{equation}
They are annihilated by all the generators  $L_n$ with $n > 0 $
\begin{equation}
L_n \vert \Delta , \bar \Delta \rangle \ = \ \bar L_n \vert \Delta , \bar \Delta \rangle \
=  \  0  \qquad {\rm for } \ \ n>0 \ .
\label{eq:1.35d}
\end{equation}
The conformal dimension of a primary field is equal to the sum of its two 
conformal weights, while its spin (or helicity) is equal to their difference
\begin{equation}
d  = \Delta + \bar \Delta  \qquad \qquad s = \Delta - \bar \Delta \ .
\label{eq:1.36}
\end{equation}
There exist also fields that satisfy (\ref{eq:1.35a},\ref{eq:1.35b}) only
for $n=0,\pm1$. Such fields are called quasiprimary (or conformal
descendants). The corresponding quasiprimary states 
\index{state ! quasiprimary}
are obtained from 
the primary  states \eref{eq:1.35c} by the action of polynomials 
in $L_n$ with negative $n$. All the properties of the quasiprimary fields
follow from those of the underlying primary one.
  
\subsection{Rational conformal field theories}  

One important class of theories are the  Rational Conformal Field Theories (RCFT).
In a RCFT there are only a finite number of primary fields. For example,
in the unitary minimal models  \cite{[10],[11]} corresponding to  central charge 
of the Virasoro algebra
\begin{equation}
c = 1 - {6 \over m (m+1)} \qquad  m \geq 3 \ 
\label{eq:1.37a}
\end{equation}
the primary fields have weights  
\begin{equation}
\Delta_{r,s} = { [r(m+1) -s m ]^2 -1 \over 4 m (m+1)} \   \qquad
1 \leq r \leq m-1 \  \qquad 1 \leq s \leq m \ .
\label{eq:1.37b}
\end{equation}
Another important example are the superconformal models. 
The supersymmetry generator $G(z)$ has conformal weight $3/2$
and hence a Laurent expansion
\begin{equation}
G(z) \ = \ \sum_r {G_r \over z^{r+{3 \over 2}}} \ . 
\label{eq:1.38}
\end{equation}
Since $G(z)$ has half integer spin, it can be chosen either periodic 
(Ramond sector) or antiperiodic (Neveu-Schwarz sector) \cite{[13]}.
In the Ramond sector the sum in \eref{eq:1.38} is over $r$ integer,
while in the Neveu-Schwarz sector it is over $r$  half-integer. 
The (anti)commutation relations between $L_n$ and $G_r$ are 
\begin{eqnarray}
\left[ L_n , G_r \right] \ &=&  \ \left( {n \over 2} - r \right) G_{n+r} \   
\label{eq:1.39a} \\
\ms
\left\{ G_r , G_s \right\} \ &=& \ 2 L_{r+s} + {c \over 3} \left(r^2 - {1 \over 4} \right)
\delta_{r+s} \ . 
\label{eq:1.39b}
\end{eqnarray}
The unitary $N=1$ superconformal models   have central charge
\begin{equation}
c = {3 \over 2} \left [1 - {8 \over m (m+2)} \right]
\qquad  m \geq 3 \ ,
\label{eq:1.40a}
\end{equation}
while the conformal weights of the primary fields are \cite{[14],[15]}
\begin{equation}
\Delta_{r,s} = { [r(m+2) -s m ]^2 -4 \over 8 m (m+2)} 
\ + \ {1 \over 32} [1- (-1)^{r-s} ] \  , 
\label{eq:1.40b}
\end{equation}
where $1 \leq r \leq m-1$ and $1 \leq s \leq m $. The
 Neveu-Schwarz sector contains the fields with  $r-s$ even, while the  Ramond
sector contains the fields with $r-s$ odd.

In order to describe the $N=2$ superconformal models \cite{[16]} it is convenient 
to  study  first the simplest example of a conformal current algebra, namely the 
abelian $U(1)$ case. The chiral part of the $U(1)$ current satisfying  \eref{eq:1.21}
has the following expansion in Laurent modes
\begin{equation}
J(z) \ = \  \sum_{n}{J_n \over z^{n+1}} \quad \qquad
J_n^{\dagger} = J_{-n} \ . 
\label{eq:1.41}
\end{equation}
Since the $U(1)$ current is a primary field of the Virasoro algebra 
of weight one, its commutation relations 
with the modes of the stress energy tensor are
\begin{equation}
[L_n,J_m]=-m\ J_{m+n}\ .  
\label{eq:1.42}
\end{equation}
The locality condition \eref{eq:1.24} determines completely also the 
commutation relations between two currents 
\begin{equation}
[J(z_1), J(z_2)] = -\delta'(z_{12})  \qquad {\rm or } \qquad [J_n, J_m] =
n\delta_{n+m} \ ,
\label{eq:1.43}
\end{equation}
where for convenience we have chosen to normalize the central term to one. 
The same relations hold also for the antichiral components.
The primary fields of the $U(1)$ conformal current algebra are characterized
by their charges $q$ and $\bar q$ and satisfy the following commutation relations 
with the current components
\begin{eqnarray}
\left[ J(z_1), \phi_{ q \bar q}(z_2, \bar z_2) \right] &=& -  q \phi_{ q \bar q} (z_2, \bar
z_2)\
\delta(z_{12}) \  
\label{eq:1.44a}  \\ 
\ms
\left[ \bar J( \bar z_1), \phi_{ q \bar q}(z_2, \bar
z_2) \right] &=& - \bar q \phi_{ q \bar q}(z_2, \bar z_2)\ \delta(\bar z_{12}) \ .
\label{eq:1.44b}
\end{eqnarray}
The stress energy tensor can be expressed in terms of the currents by the 
Sugawara formula \cite{[38]} \index{Sugawara formula}
 and the central charge of the Virasoro algebra is equal to one
\begin{equation}
T(z)={1\over 2}:J^2(z): \qquad \Rightarrow \quad c(u(1))=1\ , 
\label{eq:1.45a}
\end{equation}
which for the Laurent modes gives
\begin{equation}
L_n={1\over 2}\left(\sum_{m\geq 1}+\sum_{m\geq -n}\right)\ J_{-m}\ J_{m+n} \ .
\label{eq:1.45b}
\end{equation}
The consistency of equations 
 \eref{eq:1.44a}, \eref{eq:1.45b} and  \eref{eq:1.35a} implies a relation 
between the $U(1)$ charges and the conformal weights  
\begin{equation}
\Delta = {1 \over 2} q^2 \  \qquad \bar \Delta = {1 \over 2} \bar q^2 \ ,
\label{eq:1.46}
\end{equation}
as well as the following  equations for the primary fields \cite{[17],[18]}
\begin{eqnarray}
&& \partial_{z} \phi_{ q \bar q}(z, \bar z) + q : J(z)\ \phi_{ q \bar q}(z, \bar
z) :\  = \ 0 \   
\label{eq:1.47a} \\
\ms
&& \partial_{\bar z} \phi_{ q \bar q}(z, \bar z) + \bar q : \bar J(\bar z)\ \phi_{
q \bar q}(z,
\bar z) :\  = \ 0 \ . 
\label{eq:1.47b} 
\end{eqnarray}

The $N=2$ superconformal algebra contains two supersymmetry 
generators  $G^{\alpha}(z)$, $\alpha = 1,2$,  with Laurent expansions 
\eref{eq:1.38} and a $U(1)$ current $J(z)$ with expansion \eref{eq:1.41}.
The new (anti)commutation relations are 
\begin{eqnarray}
\left\{ G_r^{\alpha} , G_s^{\beta} \right\} \ & = & \ 2 
\delta ^{\alpha \beta} L_{r+s} \nonumber \\
&& + \rmi (r - s) \epsilon^{\alpha \beta} J_{r+s}
+ {  c  \over 3}
\left( r^2 - {1 \over 4} \right) \delta ^{\alpha \beta}
\delta_{r+s} \  
\label{eq:1.48a} \\
\ms
\left[ J_m , G_r^{\alpha} \right] \ & = & \rmi \epsilon^{\alpha \beta}
G_r^{\beta} \ , 
\label{eq:1.48b} 
\end{eqnarray}
where $\epsilon^{\alpha \beta}$ is antisymmetric and  $\epsilon^{12}=1$.
There are three sectors: in the Neveu--Schwarz and  Ramond sectors the 
$U(1)$ current has integer modes, while in the twisted sector the 
$U(1)$ current has  half integer modes \cite{[19]}.
The unitary minimal  $N=2$ superconformal models  correspond to central charges 
\begin{equation}
 c = 3 \left( 1 -{2 \over m}\right) \   \qquad  m \geq 3 \quad . 
\label{eq:1.49} 
\end{equation}

\subsection{Nonabelian conformal current algebras}

The nonabelian generalization of the $U(1)$ conformal current algebra  \eref{eq:1.43}
known also as Wess-Zumino-Witten (WZW) model is one of the few cases  
of two dimensional CFT for which one can write also an explicit action \cite{[33]}.
Alternatively one can use the following definition.
Let $G$ be a compact semi-simple Lie group and $\cal G$ be its 
Lie algebra of dimension $d_G$. 
The chiral conformal current algebra 
 ${\cal A}({\cal G})$
is the algebra generated by the  $d_G$ chiral currents 
in  the adjoint representation of $\cal G$.
The  currents are primary fields of the Virasoro algebra of conformal weight one
and have the  Laurent expansion
\begin{equation}
J^a(z) = \sum_{n}\ {J^a_n \over z^{n+1}} \quad \qquad J^{a^*}_n=J^a_{-n} \ .
\label{eq:2.15a}
\end{equation}
The commutation relations for their modes are \index{algebra ! affine Kac-Moody}
\begin{equation}
[J^a_n,J^b_m] = {\rm i} \sum_c \ f_{abc}\ J^c_{n+m}+{k \over 2} n\ \delta_{ab}\
\delta_{n+m}
\ , \label{eq:2.16a}
\end{equation}
where $f_{abc}$  are the structure constants of  $\cal G$
and the level $k$ is a nonnegative integer.
These relations define an affine Kac-Moody algebra \cite{[12]}.

The stress energy tensor can be expressed 
in terms of the currents \eref{eq:2.15a}  by the Sugawara formula 
\index{Sugawara formula}
\begin{equation}
2h\ T(z) \ = \ \sum^{d_G}_{a=1}:J^2_a(z):\ 
 , \label{eq:2.19}
\end{equation}
where the height $h$ is  the sum of the level $k$ and the dual 
Coxeter number of $\cal G$, $h=k+g\check{}$ (= $k+N$ for $SU(N)$). 
In terms of the  Laurent modes  \eref{eq:2.19}  becomes
\begin{equation}
2hL_n =\left(\sum^\infty_{\ell =1} +\sum^\infty_{\ell =-n}\right)\
\sum^{d_G}_{a=1} 
 J_{-\ell}^a \  J_{n+\ell}^a \ , \label{eq:2.21c}
 \end{equation}
while the central charge of the  Virasoro algebra is  
\begin{equation}
c \ = \ {k\over h}\ d_G \ . \label{eq:2.22}
\end{equation}

The primary fields  of ${\cal A}({\cal G})$ \index{field ! primary}
are in one-to-one correspondence with the irreducible 
representations of ${\cal G}$, hence we can label them by  
highest weight vectors $\Lambda =(\lambda_1,\dots ,\lambda_r)$ of $\cal G$.
We shall denote the primary fields by $V_\Lambda (z)$. 
They satisfy the following commutation relations with the currents
(for brevity we  omit the dependence on $\bar z$ and  write only the
relations in the  chiral sector)   
\begin{equation}
[J^a(z_1),V_\Lambda
(z_2)]=\delta (z_{12})\ V_\Lambda (z_2)\ t_\Lambda^a \  \label{eq:2.37a}
\end{equation}
or in terms of the modes \eref{eq:2.15a}
\begin{equation}
[J^a_n, V_\Lambda (z)]=z^n\ V_\Lambda (z)\
t_\Lambda^a \ , \label{eq:2.37b}
\end{equation}
where  $t_\Lambda^a$ are the matrices of  $J^a_0$ in the representation 
$\Lambda$.
The consistency of   equations \eref{eq:2.21c} and \eref{eq:2.37b} with \eref{eq:1.35a}
implies the relation 
\begin{equation}
2h\ \Delta_\Lambda =C_2(\Lambda ) \  
\label{eq:2.30}
\end{equation}
between the conformal weight of the primary field and the eigenvalue of the 
second order Casimir operator in the representation  $\Lambda $, as well as 
the operator form of the Knizhnik--Zamolodchikov  (KZ) equation \cite{[17],[18]} 
\index{KZ equation}
\begin{equation}
h\ {d\over dz}\ V_\Lambda (z)=\sum_{a=1}^{d_G} :V_\Lambda (z)\ t^a_\Lambda\ J^a(z): \
.
\label{eq:2.102}
\end{equation}

The primary  fields in a two dimensional conformal theory
transforming as in \eref{eq:1.34} in general do not split in a sum of 
 chiral and antichiral components.
Rather they are given  by a (finite in the case of 
a rational conformal theory) sum of products of
chiral and antichiral vertex operators  \cite{[39],[61]}.
In order to properly define a 
 chiral vertex operator \index{chiral vertex operator} we have to specify  
 a triple of weights $\left({\Lambda_f\atop\Lambda \ \ \Lambda_i}\right)$
where  
$\Lambda_i$ is the weight on which  $V_\Lambda$ acts,
while   $\Lambda_f$ is the weight to which $V_\Lambda$ maps.
In other words, the chiral vertex operators can be represented as 
\begin{equation}
V\kern-5pt\textstyle{{\Lambda_f\atop\Lambda\ \  \Lambda_i}} (z) =\Pi_{\Lambda_f}\
V_\Lambda\ (z) \
\Pi_{\Lambda_i} \ , \label{eq:2.42}
\end{equation}
where $\Pi_\Lambda$ are orthogonal projectors,
and  in general are multivalued functions of $z$
\begin{equation}
V\kern-5pt\textstyle{{\Lambda_f\atop\Lambda \ \ \Lambda_i}} (\rme^{2\pi
 \rmi}z)=\rme^{
2\pi \rmi (\Delta_{\Lambda_f}-\Delta_\Lambda -\Delta_{\Lambda_i})}
V\kern-5pt\textstyle{{\Lambda_f\atop\Lambda\ \ \Lambda_i}} (z) \ .
\label{eq:2.41}
\end{equation}
The correlation functions of the chiral vertex operators are called 
chiral conformal blocks and due to \eref{eq:2.41} are also multivalued 
functions of the coordinates.
The  two dimensional primary fields $\phi (z,\bar z)$  \index{field ! bulk}
can be written 
in terms of the chiral vertex operators \eref{eq:2.42} 
as 
\begin{equation}
\phi_{\Lambda\bar\Lambda}(z,\bar z) =\sum_{{\Lambda_i\ \ \bar\Lambda_i\atop
\Lambda_f\ \ \bar\Lambda_f}}\ V\kern-5pt\textstyle{{\Lambda_f\atop\Lambda\ \
\Lambda_i}}(z)\
\bar V\kern-5pt\textstyle{{\bar\Lambda_f\atop\bar\Lambda\ \ \bar\Lambda_i}}(\bar
z) \ .
\label{eq:2.45}
\end{equation}
Locality and \eref{eq:2.41} imply that the  
spin of all fields $\Delta_\Lambda -\Delta_{\bar\Lambda}$
has to be integer.
Note that this selection rule must be respected also by the 
pairs of weights 
$(\Lambda_i,\bar\Lambda_i)$ and $(\Lambda_f,\bar\Lambda_f)$. 
One large class of theories which satisfy trivially this requirement are the diagonal 
theories  with  $\Lambda =\bar\Lambda$.

\subsection{Partition function, modular invariance}

Due to the factorization of the observable algebra \eref{eq:1.22} 
we can analyze independently the chiral and antichiral sectors, 
but in order to reconstruct the whole two dimensional theory 
we need  also the pairings between the fields from the two sectors.
They can be found   requiring the modular invariance of the partition function 
on the torus. From the viewpoint of string theory  the modular invariance condition is 
very natural, since it ensures that one can define the theory on surfaces 
of arbitrary genus \cite{[21],[Nahm]}. In Statistical Mechanics models its physical meaning 
is more subtle, since the modular transformations relate the low and the 
high temperature behaviour of the theory \cite{[22]}.

Let us briefly recall the construction of the partition function.
To every primary field $\varphi_i$ of  $\cal A$ corresponds a character 
\index{character}
of the Virasoro algebra \cite{[12]}
\begin{equation}
\chi_i(\tau) \ = \ Tr_{{\cal H}_i} e^{2 \pi i \tau (L_0 - {c \over 24})} \ ,
\label{eq:1.50} 
\end{equation} 
where the trace is over the space of 
all  quasiprimary descendants of  $\varphi_i$. Note that the 
energy operator $L_0$ on the torus is modified according to 
\eref{eq:1.27}.
In this notation the torus partition function
\index{partition function ! torus}
\begin{equation}
Z_T \ = \ Tr \left( e^{2 \pi i \tau (L_0 - {c \over 24})} \ 
e^{ 2 \pi i \bar \tau (\bar L_0 - {\bar c \over 24})} \right) \ 
\label{eq:1.51a} 
\end{equation}
can be rewritten as  (we recall that $\bar c = c$)
\begin{equation}
Z_T \ = \ \sum_{i,j} \chi_i \  X_{i j} \ \bar \chi_j \ ,
\label{eq:1.51b} 
\end{equation}
where  $X_{ij}$ are non-negative integers which give the 
multiplicities of the two dimensional fields. 
For the rational theories the sum in 
 \eref{eq:1.51b} is over a finite set of characters.

Not all values of $\tau$ in \eref{eq:1.51a} correspond to 
inequivalent tori. In particular the transformations  
\begin{eqnarray}
S \ &:& \ \tau \longrightarrow - {1 \over \tau} \  \label{eq:1.52a} \\ 
T \ &:& \ \tau \longrightarrow { \tau + 1 } \  \label{eq:1.52b} 
\end{eqnarray}
are just redefinitions of the fundamental cell of the torus.
They generate the modular group  $PSL(2,\mathbb Z)$ under which $\tau$
transforms as \index{modular transformations}
\begin{equation}
\tau \  \longrightarrow \ \tau^{'} \ = \ 
{a \tau  + b \over c \tau +d} \   \qquad 
ad-bc = 1 \   
\label{eq:1.53} 
\end{equation}
with integer $a,b,c$ and  $d$.
These transformations act linearly on the characters
 \eref{eq:1.50}
\begin{equation}
\chi_i \left(- \ {1 \over \tau}\right) \ = \ \sum_j \ S_{ij} \ \chi_j(\tau)
\ 
\qquad
\chi_i (\tau +1) \ = \ \sum_j \ T_{ij} \ \chi_j(\tau)  \ ,
\label{eq:1.54} 
\end{equation}
where  $T$ is a diagonal matrix, while  $S$ is a symmetric matrix.
Both  $S$ and $T$ are unitary and satisfy 
$S^2 = (ST)^3 = C$, where the matrix $C$ is 
called charge conjugation matrix and satisfies $C^2=1$. 

The modular invariance of the torus partition function implies 
\begin{equation}
  S X S^{\dagger} = X 
\qquad  \qquad  T X T^{\dagger} = X \ . 
\label{eq:1.55} 
\end{equation}
The solutions to these equations are of two distinct types \cite{[40]}. 
The first one are   called permutation (or automorphism) 
invariants, for which 
\begin{equation}
X_{ij} = \delta_{i \sigma (j)} \quad , 
\label{eq:1.56} 
\end{equation}
where $\sigma (j)$ is a permutation of the labels $j$.
The second one correspond to extensions of the observable algebra
and  can always be  rewritten as a
permutation invariant \eref{eq:1.56} in terms of the characters of the 
maximally extended observable algebra (that are linear combinations of 
the characters of the unextended one). 

Let us denote by $[\varphi_i]$ the conformal family of the primary field 
$\varphi_i$ i.e. the collection of all the conformal descendants of $\varphi_i$.
The product of two  conformal families is determined by the fusion algebra 
\begin{equation}
[\varphi_i] \times [\varphi_j] \ = \ \sum_k \ {N_{ij}}^k \ [\varphi_k] \quad .
\label{eq:1.57} 
\end{equation}
The non negative integers 
 ${N_{ij}}^k$, called fusion rules 
 can be expressed in terms of the modular matrix $S$  by the Verlinde formula 
\begin{equation}
{N_{ij}}^k \ = \ \sum_{\ell} \ { S_{i \ell} S_{j \ell} S_{k \ell}^{\dagger} 
\over S_{1 \ell}}  \   
\label{eq:1.59} 
\end{equation}
and  as matrices ${(N_i)_j}^k$ satisfy the commutative and associative fusion algebra 
\index{algebra ! fusion} \cite{[23]}
\begin{equation}
(N_i) \ (N_j) \ = \sum_k {N_{i j}}^k (N_k) \quad . 
\label{eq:1.58} 
\end{equation}

There are several known classifications of 
modular invariant partition functions, 
e.g.\cite{[24],[26],[31],[32]}, but  the problem 
is still not solved in general. We shall often refer to the
$A-D-E$ classification of Cappelli, Itzykson and Zuber \cite{[24]} of the
modular invariants of the $SU(2)$ conformal current algebra. 
In this classification, the diagonal $ \ A \ $ and 
the $D_{odd}$ series are permutation invariants, the 
$D_{even}$ series,  $E_6$ and $E_8$ are diagonal invariants
of an extended algebra, while $E_7$ is a nontrivial 
permutation invariant of an extended algebra.  

There is also an alternative method to compute the allowed 
pairings between the fields of the two sectors that makes no use
of higher genus partition functions.
In two dimensional conformal field theory the product of two 
primary fields can be expressed as a 
sum of primary fields and their conformal descendants
using the Operator Product Expansion (OPE) \index{operator product expansion ! bulk}
\begin{eqnarray}
&&\phi_{\Delta_i, \bar \Delta_i}(z,\bar z)
\phi_{\Delta_j, \bar \Delta_j}(w,\bar w) \ \nonumber \\
\ms
&&  = \
\sum_{k, \bar k}
{C_{(i, \bar i)(j, \bar j)}^{(k, \bar k)} 
\over (z-w)^{\Delta_i+\Delta_j-\Delta_k} 
(\bar z-\bar w)^{\bar \Delta_i+\bar \Delta_j-\bar \Delta_k}}
\phi_{\Delta_k, {\bar \Delta}_k}(w,\bar w) + \dots
\label{eq:3.21blabla}
\end{eqnarray}
where the dots stand for the descendants.
The two dimensional structure constants \index{structure constants ! bulk}
$C_{(i, \bar i)(j, \bar j)}^{(k, \bar k)}$  
vanish whenever the corresponding fusion rules 
${N_{ij}}^k$  or ${N_{\bar i \bar j}}^{\bar k}$ are zero and 
completely define the theory. 
In particular they determine also the allowed 
pairings between the fields of the two sectors. 
Moreover they  
permit to reconstruct all the Green functions of the two dimensional 
fields. In rational conformal field theories the 
structure constants can in principle be computed  imposing the 
 locality (or crossing symmetry)
of the 4-point Green functions.
Indeed for a generic 4-point function 
\begin{equation}
\langle 
\phi_{\Delta_1, \bar \Delta_1}(z_1,\bar z_1) \
\phi_{\Delta_2, \bar \Delta_2}(z_2,\bar z_2) \
\phi_{\Delta_3, \bar \Delta_3}(z_3,\bar z_3) \
\phi_{\Delta_4, \bar \Delta_4}(z_4,\bar z_4)
\rangle \nonumber
\end{equation}
we can apply the OPE \eref{eq:3.21blabla}
in three different ways which schematically 
can be denoted as $(12)(34)$, $(13)(24)$
and $(14)(23)$. This gives two duality relations 
between the structure constants and determines
them up to global rescalings of the two dimensional
fields. In practice this procedure is very complicated
and the closed expressions for the  two dimensional structure constants
are known only in a very limited number of cases (in particular for
 the $SU(2)$ current algebra models and for  the 
unitary minimal models \cite{[64],[66]}).

Let us stress that while the crossing symmetry 
relations are satisfied also for any  
subset of primary fields closed under OPE, e.g.\ for the identity operator alone 
to give a trivial example, the modular invariance condition
is satisfied only by the maximal (or complete) set of fields. 

In fact these two approaches are complementary, since
as demonstrated in \cite{[62],[61]} both the condition
of crossing symmetry of the 4-point functions and the modular invariance 
of the torus partition function are necessary and sufficient 
for the consistency of the theory
on a surface of arbitrary genus.


\section{Correlation functions in current algebra \\ models}
\label{sec3}

In the conformal current algebra models
the operator Knizhnik-Zamolod\-chikov equation 
\eref{eq:2.102} implies a system of first order
partial differential equations for the $n$-point chiral conformal blocks. 
This allows one to reformulate all the properties of the primary conformal fields 
as conditions on their chiral correlators.
Moreover, for the $SU(2)$ models that  
we shall review in some detail 
this also allows  to obtain explicit expressions 
for the chiral conformal blocks and to compute the structure constants that 
enter the two dimensional 
operator product expansion \eref{eq:3.21blabla}.

\subsection{Properties of the chiral conformal blocks}

Let $G$ be a simply  connected compact Lie group  with Lie algebra 
${\cal G}$ and let $V_i= V(\Lambda_i)$, $i=1,2,\dots, n$ be
chiral vertex operators of highest weight  
$\Lambda_i$ such that the space ${\cal J}_n={\cal J}(\Lambda_1,\dots \Lambda_n)$ 
of $G$ invariant tensors is 
non trivial ($d_{\cal J} = \dim {\cal J}_n >0$).
Consider the $d_{\cal J}$ dimensional vector space ${\cal L}_n$
of holomorphic functions $w_n=w(z_1,\Lambda_1;\dots; 
z_n,\Lambda_n)$ called chiral conformal blocks \cite{[10]}
with values in ${\cal J}_n$.

M\"obius invariance of the vacuum implies that the functions  $w_n$ 
are covariant under local M\"obius transformations.
In particular they are translation invariant (hence depend only on the differences
$z_{ij}$), they transform covariantly under uniform dilations $z_i\to
\rho z_i$, $\rho>0$
\begin{equation}
\rho^{\Delta_1+\dots +\Delta_n} w(\rho z_1, \Lambda_1;\dots;\rho
z_n,\Lambda_n)=w(z_1,\Lambda_1;\dots; z_n,\Lambda_n) \ , 
\label{eq:3.3a}
\end{equation}
where  
$\Delta_i=\Delta(\Lambda_i)$ are the conformal weights \eref{eq:2.30}.
Finally,  $w_n$ are covariant under infinitesimal special conformal
transformations $z\to z/(1+\varepsilon z)$ with $\varepsilon \to 0$,
thus satisfy the differential equation
\begin{equation}
\sum^n_{i=1}\  z_i \left(z_i{\partial\over \partial z_i}
+2\Delta_i\right) w_n =
0\ .
\label{eq:3.4}
\end{equation}

The operator form of the  Knizhnik-Zamolodchikov  equation \eref{eq:2.102} implies that 
all elements in ${\cal L}_n$ satisfy the system of partial differential equations \cite{[17]}
\index{KZ equation}
\begin{equation}
\left({\partial\over\partial z_i} +{1 \over h} 
\sum^n_{j=1\atop j\not= i} { \sum_{a} t^a_{\Lambda_i} t^a_{\Lambda_j} 
\over z_{ij}} \right) w_n=0 \ 
\label{eq:3.5a}
\end{equation}
for $i=1,\dots,n$, where $h$ is the height defined after equation 
 \eref{eq:2.19}.

Every function  $w_n$ of ${\cal L}_n$ admits a path dependent multivalued
analytic continuation in the product of complex planes minus the diagonal
$\{z_i\in {\mathbb C},\ z_i\not= z_j\
{\rm for}\ i\not= j\} $.
Let us choose a basis $\{w^\nu_n,\ \nu =1,\dots, d_{\cal J}\}$
in ${\cal L}_n$ and consider the analytic continuation of $w^\nu_n$ 
along a pair of paths ${\cal C}^\pm_i$ that exchange two 
neighbouring arguments $z_i$, $z_{i+1}$ in positive/negative
directions
\begin{equation}
{\cal C}^\pm_i:\left({z_i\atop z_{i+1}}\right)\to {1\over
2}(z_i+z_{i+1})+{1\over
2}
\left({z_{ii+1}\atop -z_{ii+1}}\right) e^{\pm i\pi t} \ ,
\label{eq:3.10}
\end{equation}
where $0\leq t\leq 1 $.
This operation followed by the permutation of the two weights  $\Lambda_i$ a $\Lambda_{i+1}$ 
defines the action of  two exchange operators 
 $B_i$ and $\bar B_i$ \cite{[39],[61],[60]}.
The exchange operator  $B_i$ transforms the basis 
$\{ w^\nu_n\}$ in ${\cal L}(\Lambda_1,\dots , \Lambda_i,
\Lambda_{i+1},\dots,\Lambda_n)$ in  a basis 
$\{w^\mu_n\}$ in ${\cal L}(\Lambda_1,\dots,\Lambda_{i+1},
\Lambda_i,\dots,\Lambda_n)$. 
\begin{eqnarray}
B_i=B^{\Lambda_1\dots \Lambda_n}_i \ :
&& {\cal L}(\Lambda_1,\dots,\Lambda_i,\
\Lambda_{i+1},\dots,\Lambda_n) \nonumber \\
\ms
&& \to {\cal L}
(\Lambda_1,\dots,\Lambda_{i+1},\Lambda_i,\dots \Lambda_n)\ \ . 
\label{eq:3.11}
\end{eqnarray}
The exchange operator $\bar B_i$ is the inverse to $B_i$. More precisely 
\begin{equation}
\bar B^{\Lambda_1\dots\Lambda_{i+1}\Lambda_i\dots \Lambda_n}_i
B^{\Lambda_1\dots \Lambda_i \Lambda_{i+1}\dots \Lambda_n}_i ={\bf 1} \
.
\label{eq:3.12}
\end{equation}
 For real analytic  $w^\nu_n$ the matrix
$\bar B_i$ is complex conjugate to  $B_i$.
The operators $B_i$,
$i=1,\dots,n-1$ with various order of the weights $(\Lambda_1,\dots,\Lambda_n)$
generate a  representation of the exchange (called also braid \cite{[59]}) algebra ${\cal B}_n$.

 The two dimensional 
$n$-point Green functions $G_n$
 can be  written as a finite sum of products of  $n$-point 
chiral and antichiral blocks \begin{eqnarray}
G_n &=& \langle 0\vert \phi_1( z_1, \bar z_1)\dots\phi_n( z_n,\bar z_n)\vert 0\rangle
\nonumber \\
\bs
& = & \bar w^\mu_n \ Q^{\Lambda_1\dots\Lambda_n}_{\mu\nu}  \  w^\nu_n \ .
\label{eq:3.18a}
\end{eqnarray}
Local commutativity of the two dimensional
fields is equivalent to the invariance of the Green functions $G_n$ 
under the combined action of the two exchange algebras
which implies 
a braid invariance condition for 
the matrices $Q^{\Lambda_1\dots\Lambda_n}$ \cite{[60]}
\begin{equation}
(B^{\Lambda_1\dots\Lambda_i\Lambda_{i+1}\dots\Lambda_n}_i)^{\dagger}
Q^{\Lambda_1\dots\Lambda_i \Lambda_{i+1}\dots\Lambda_n}
B^{\Lambda_1\dots\Lambda_{i+1}\Lambda_i\dots\Lambda_n}_i =
Q^{\Lambda_1\dots \Lambda_{i+1}\Lambda_i\dots\Lambda_n} \ .
\label{eq:3.19}
\end{equation}
The relative normalization of $G_n$ for different $n$ and different 
sets of weights are constrained by the  
factorization 
properties implied by the two dimensional operator product expansion
\eref{eq:3.21blabla}.

\subsection{Regular basis of 4-point functions in the $SU(2)$ model}

We shall consider in some detail only 
the simplest non-trivial case  of 4-point functions for $G=SU(2)$. 
Note that there is an infinite series of such models 
 corresponding to integer height $h=k+2$ and Virasoro central charge
$c={3 k  \over k+2}$. The primary fields can be labelled by their
isospin $I$ which has to satisfy the integrability condition 
 $I \leq k/2$ \cite{[41]} and have conformal dimension
$\Delta(I)={I(I+1) \over (k+2)}$. 
The fusion rules can be computed from the Verlinde formula  \eref{eq:1.59}
and in terms of the isospins of the fields are 
\begin{equation}
[I_1]\times [I_2]=\sum^{{\rm min}(I_1+I_2,k-I_1-I_2)}_{I=|I_1-I_2|}\ [I] \ . \label{eq:2.56a}
\end{equation}

Exploiting M\"obius invariance one can reduce the  KZ equation \eref{eq:3.5a}
to a system of ordinary differential equations. In order to write more compact 
formulae we shall make use of the polynomial realization of the irreducible
$SU(2)$ modules \cite{[50]} and  
introduce an auxiliary variable $\zeta$ to keep track of the third
isospin projection $m$ of the operators. In particular we shall set
\begin{equation}
V_{I}(z,\zeta) \ = \ \sum_{m=-I}^{m=I}
{\zeta^{I+m} \over (I+m)!} V_{I}^m(z) \ .
\label{eq:3.zeta}
\end{equation}
The $SU(2)$ generators act on $V_{I}(z,\zeta)$ as first 
order differential operators in $\zeta$, while the 
correlation functions are polynomials in $\zeta$.
We shall  also assume that 
the isospins of the four fields 
satisfy the inequalities ($I_{ij}=I_i-I_j$)
\begin{equation}
I(=\min I_i) =I_4 \ \qquad \vert I_{12}\vert\leq I_{34} \ \qquad \vert
I_{23}\vert\leq I_{14} \ . 
\label{eq:3.30}
\end{equation}
The other cases can be treated in exactly the same way.

M\"obius and $SU(2)$ invariance imply that the 4-point chiral conformal blocks 
have the form 
\begin{equation}
w(z_1, \zeta_1, I_1; \dots; z_4,\zeta_4,
I_4)=g(z_{ij},\Delta)p(\zeta_{ij},I_{ij})F(\eta,\xi_1,\xi_2)\ . 
\label{eq:3.25a}
\end{equation}
Here  $g(z_{ij},\Delta)$ is a scale prefactor
\begin{equation}
g(z_{ij},\Delta)=
{z^{\Delta_2+\Delta_4}_{13} z^{\Delta_1+\Delta_3}_{24}
\eta^{\Delta_s}(1-\eta)^{\Delta_u}\over
z^{\Delta_1+\Delta_2}_{12} z^{\Delta_2+\Delta_3}_{23} z^{\Delta_2+\Delta_4}_{34}
z^{\Delta_1+\Delta_4}_{14}} \ ,
\label{eq:3.25b}
\end{equation}
 $\eta$ is the M\"obius  invariant crossratio
\begin{equation}
\eta={z_{12}z_{34}\over z_{13}z_{24}} \
\left(=1-{z_{14}z_{23}\over z_{13}z_{24}}\right)\ ,
\label{eq:3.29}
\end{equation}
while $\Delta_s$ and $\Delta_u$ are the threshold dimensions in the 
 $s-$ $(12)(34)$ and  $u-$ $(23)(14)$ channels. For isospins constrained by 
 \eref{eq:3.30}  they are given by 
\begin{equation}
\Delta_s=\Delta(I_{34})={1\over h}
I_{34}(I_{34}+1) \  \qquad 
\Delta_u = \Delta(I_{14})={1\over h} I_{14}(I_{14}+1)\ .
\label{eq:3.32}
\end{equation}
The 
polynomial $p( \zeta_{ij}, I_{ij})$  is 
\begin{equation}
p( \zeta_{ij}, I_{ij}) = \zeta^{I_{14}+I_{23}}_{12} \zeta^{I_{34}-I_{12}}_{23}
\zeta^{I_{12}+I_{34}}_{13} \ . 
\label{eq:3.31}
\end{equation}
Finally, the  M\"obius invariant function $F$ 
is a homogeneous polynomial  
\begin{equation}
F(\eta;\xi_1,\xi_2)=\sum^{2I}_{\ell=0}(\xi_2\eta)^\ell
[\xi_1(1-\eta)]^{2I-\ell}f_\ell(\eta)\ 
\label{eq:3.28}
\end{equation}
in the combinations 
\begin{equation}
\xi_1=\zeta_{12}\zeta_{34},\quad  \xi_2=\zeta_{14}
\zeta_{23} \ , \quad (\xi_1+\xi_2=\zeta_{13}\zeta_{24}) \ .
\label{eq:3.27}
\end{equation}

Inserting these formulae into the KZ  equation \eref{eq:3.5a} for  
$n=4$
after some algebra we obtain a 
system of first order ordinary differential equations for the functions $f_\ell(\eta)$
\begin{eqnarray}
{df_\ell\over d\eta} &=&\left\{{\ell\over
\eta}[\alpha+\gamma-1+(\ell-1)\delta]-{2I-\ell\over
1-\eta}[\beta+\gamma-1\right. \nonumber \\
&+&(2I-\ell-1)\delta]\biggr\}f_\ell +{\ell +1\over 1-\eta}(\alpha+\ell
\delta)f_{\ell+1}- \nonumber \\ 
&-&{2I-\ell+1\over \eta}[\beta+(2I-\ell)\delta]f_{\ell-1}\ ,
\label{eq:3.41}
\end{eqnarray} 
where
\begin{eqnarray}
h\alpha =1+I_{34}-I_{12}\ &,& \quad h\beta=1+I_{14}+I_{23}\ ,\quad
h\gamma=1+I_{34}+I_{12} \nonumber \\
&& h\delta=1 \ , \qquad (h=k+2)\ .
\label{eq:3.42}
\end{eqnarray}

The system  \eref{eq:3.41} has 
 $2I+1$ linearly independent solutions 
$f_{\lambda\ell}$,
$\lambda=0,1,\dots, 2I$  which for  
$0<\eta<1$ are given by the integral representations \cite{[45],[63]}
\begin{eqnarray}
f_{\lambda\ell}(\eta) &=&  \int^\eta_0 dt_1 \int^{t_1}_0 dt_2\dots
\int^{t_{\lambda-1}}_0 \!\!\!\!\! dt_\lambda
\int^1_\eta dt_{\lambda+1}  \nonumber \\
\ms
&& \times 
\int^1_{t_{\lambda+1}} \!\!\!\!\!
 dt_{\lambda+2}\dots \int^1_{t_{2I-1}} \!\!\!\!\! dt_{2I} P_{\lambda\ell}
(t_i;\eta;\alpha,\beta,\gamma,\delta) \ ,
\label{eq:3.43}
\end{eqnarray}
where
\begin{eqnarray}
P_{\lambda\ell}&=&\mathop{\Pi}\limits^{2I}_{i=1}
t^\alpha_i(1-t_i)^\beta
\mathop{\Pi}\limits^\lambda_{i=1}
(\eta-t_i)^{\gamma-1}
\mathop{\Pi}\limits^{2I}_{j=\lambda+1}
(t_j-\eta)^{\gamma-1}
\mathop{\Pi}\limits_{i<j} (\varepsilon_{\lambda j} t_{ij})^{2\delta} \nonumber \\
\ms 
&& \times \sum_\sigma {1\over \ell!(2I-\ell)!}
\mathop{\Pi}\limits^\ell_{s=1} t^{-1}_{i_s} \mathop{\Pi}\limits^{2I}_{r=\ell+1}
(1-t_{i_r})^{-1}\ , \label{eq:3.44a} \\
\bs
\varepsilon_{\lambda j}&=&\left\{{1\ {\rm for}\ \lambda\geq j\atop
 -1\ {\rm for}\ \lambda <j}\ ,\right.\quad
t_{ij}=t_i-t_j\ .
\nonumber
\end{eqnarray}
The sum in  \eref{eq:3.44a} extends over all  $(2I)!$ permutations 
$\sigma : (1,\dots, 2I)\to (i_1,\dots, i_{2I})$. 
Note that the integration contours in  \eref{eq:3.43} never go to infinity.
This is an important  difference with respect to the commonly used integral
representations \cite{[50],[64],[66]} which correspond to tree expansions.
Our choice has the advantage that the solutions are 
linearly independent and non singular 
(if all four external isospins satisfy the integrability condition $I_i \leq k/2$). 
In particular the exchange operators are also well defined. 

\subsection{Matrix representation of the exchange algebra}

Each basis of solutions $\{w^\lambda,\ \lambda = 0,\dots, 2I\}$
of the (4-point) KZ equation gives rise to a matrix representation of the 
algebra of exchange operators $B_1$, $B_2$ and $B_3$ \cite{[60],[71]}. 
We shall work out only the action of $B_1$ and $B_2$ on the 4-point blocks
  \eref{eq:3.25a} since 
$B_3$ is proportional to  $B_1$ (see equation \eref{eq:3.61} below).
According to   \eref{eq:3.10} $B_i$ act on the cross ratio  $\eta$ \eref{eq:3.29}
 as follows
\begin{eqnarray}
&&B_1:\eta \to {\eta \rme^{\rmi\pi}\over 1-\eta} \qquad \left( =
\mathop{\lim}\limits_{t\to 1} {\eta \rme^{\rmi\pi t}\over 1+ \rmi\eta \rme^{\rmi{\pi\over
2}t} \sin
{\pi\over 2} t}\right)\  \label{eq:3.45a} \\
\bs
&&B_2:\eta \to {1\over \eta} \qquad \left(=\mathop{\lim}\limits_{t\to 1}
{\eta\cos {\pi\over 2} t-\rmi\sin {\pi\over 2}t\over \cos{\pi\over 2} t-\rmi\eta\sin
{\pi\over 2}t}\right) \ . \label{eq:3.45b}  
\end{eqnarray}
The expressions within   parentheses indicate the analytic continuation path in the $\eta$
plane, hence $B_1$ carries   $\eta$ around 0 from above, while  
$B_2$ carries  $\eta$ around 1 from below.
Note that in order to specify the domain and the target space of the exchange operators 
$B_i$ one actually has to indicate all four isospins. We shall 
use the notation
\begin{eqnarray}
B^{I_1I_2I_3I_4}_1\ &:& \ {\cal L}(I_1I_2I_3I_4)\to {\cal L}
(I_2I_1I_3I_4) \ \label{eq:3.47a} \\
\ms
B^{I_1I_2I_3I_4}_2\ &:& \ {\cal L} (I_1I_2I_3I_4)\to {\cal
L}(I_1I_3I_2I_4)\ . 
\label{eq:3.47b} 
\end{eqnarray}
The action of the exchange operators on the basis constructed in the previous subsection,
$B_i:w^\lambda \to {(B_i)^\lambda}_{\mu} w^\mu$, 
can be obtained by analytic continuation of the integral representations \eref{eq:3.43}.
Note that $B_i$ not only  transform the integrand \eref{eq:3.44a}, but also reorder 
the integration contours in \eref{eq:3.43}. The explicit expressions can be 
written in a more compact form, if one introduces  $q$-deformed numbers
\begin{equation}
 [\lambda]={q^\lambda-q^{-\lambda}\over q- q^{-1}} \ 
\label{eq:3.51b} 
\end{equation}
where
\begin{equation}
q=\rme^{\rmi\pi \delta} =\rme^{\rmi{\pi\over h}}\ (\Rightarrow q^h = -1)\ \quad 
\qquad \bar q = q^{-1} \  
\label{eq:3.50} 
\end{equation} 
and   $q$-deformed binomial coefficients
\begin{equation}
\left[{\mu\atop \lambda}\right]={[\mu]!\over
[\lambda]![\mu-\lambda]!}\ \qquad [\lambda]!=[\lambda][\lambda-1]! \ \qquad
[0]!=1 \ . 
\label{eq:3.51a} 
\end{equation}
The exchange matrix  $B_1$  is upper triangular in our basis
\begin{eqnarray}
&& {(B^{I_1I_2I_3I_4}_1)^\lambda}_\mu  = 
 (-1)^{I_1+I_2-I_{34}-\mu} \nonumber \\
 \ms
&& \ \times 
q^{(I_{34}+\mu)(I_{34}+\lambda+1)+I_{12}(\mu-\lambda)-I_1(I_1+1)-I_2(I_2+1)}
\left[{\mu\atop
\lambda}\right] \ , 
\label{eq:3.48a} 
\end{eqnarray}
while the exchange matrix $B_2$ is lower triangular and is related to $B_1$
 by a similarity transformation,
\begin{equation}
B^{I_1I_2I_3I_4}_2=F^{I_2I_3I_1I_4} B^{I_3I_2I_1I_4}_1
F^{I_1I_2I_3I_4} \ . 
\label{eq:3.48b} 
\end{equation}
The matrix  $F^{I_1I_2I_3I_4}: {\cal L} (I_1I_2I_3I_4)\to {\cal L} (I_3I_2I_1I_4)$,
called fusion matrix \cite{[61]}, 
is involutive 
\begin{equation}
F^{I_3I_2I_1I_4} F^{I_1I_2I_3I_4} \ = \ 1 \   
\label{eq:3.48c} 
\end{equation}
and in the basis \eref{eq:3.43} is represented by an antidiagonal matrix
whose elements are independent of the order of the isospins
\begin{equation}
{(F^{I_1\dots I_4})^\lambda}_\mu  = \delta^{2I-\lambda}_\mu \ .
\label{eq:3.49} 
\end{equation}
Using the expressions  \eref{eq:3.48a} and \eref{eq:3.48b} one can verify 
that the exchange operators $B_i$ satisfy   the
parameter free  Yang--Baxter equation \cite{[74]}
\begin{eqnarray}
&& B^{I_2I_3I_1I_4}_1 B^{I_2I_1I_3I_4}_2
B^{I_1I_2I_3I_4}_1=B^{I_3I_1I_2I_4}_2 B^{I_1I_3I_2I_4}_1 B^{I_1I_2I_3I_4}_2   \nonumber \\
\ms
&& = (-1)^{I_1+I_2+I_{34}} q^{I_4(I_4+1)-I_1(I_1+1)-I_2(I_2+1)-I_3(I_3+1)}
F  \ . 
\label{eq:3.53}
\end{eqnarray}
Let us note also that for the 4-point functions  
$B^{I_1I_2I_3I_4}_3$
and 
$B^{I_1I_2I_3I_4}_1$  are proportional 
\begin{eqnarray}
&& B^{I_1I_2I_3I_4}_3 = 
(-1)^{I_3+I_4-I_1-I_2} \nonumber \\
\ms
&& \times q^{\{I_1(I_1+1)+I_2(I_2+1)-I_3(I_3+1)-I_4(I_4+1)\}}
B^{I_1I_2I_3I_4}_1 \ . 
\label{eq:3.61}
\end{eqnarray}
The exchange operators for the 3-point functions are just phases, since the 
space of $SU(2)$ invariants is one dimensional in this case.
They can be obtained as a special case 
(for $I_4 = 0$) from the general expressions  \eref{eq:3.48a}, \eref{eq:3.48b}
\begin{eqnarray}
B^{I_1I_2I_3}_1     &=&
(-1)^{I_1+I_2-I_{3}}
q^{I_{3}(I_{3}+1)-I_1(I_1+1)-I_2(I_2+1)}  \label{eq:3.62a} \\
\bs
B^{I_1I_2I_3}_2    &=&
(-1)^{I_2+I_3-I_{1}}
q^{I_{1}(I_{1}+1)-I_2(I_2+1)-I_3(I_3+1)}  . 
\label{eq:3.62b}
\end{eqnarray}
The exchange operator for the 2-point function
which exists only for 
 $I_2 = I_1 $ is given by \begin{equation}
B^{I_1I_2}   =
(-1)^{2I_1}
q^{-2 I_1(I_1+1)} \ . 
\label{eq:3.63}
\end{equation}

\subsection{Two dimensional braid invariant  Green functions}

So far we have computed only the exchange operators in the chiral sector of the theory.
To compute the two dimensional Green functions 
 \eref{eq:3.18a} we need also the expressions for the antichiral sector.
To derive them let us recall that the corresponding current algebras 
are isomorphic, while the orientation of the analytic continuation 
contours \eref{eq:3.10}  are opposite in the two sectors. Thus the exchange 
operators in the antichiral sector are complex conjugate of the corresponding chiral 
ones and can be obtained from them by the substitution   
$q \rightarrow q^{-1} \ (=\bar q)$ (see equation \eref{eq:3.50}).

The locality condition for the two dimensional Green functions   \eref{eq:3.18a}
implies the braid invariance constraints \eref{eq:3.19} for the matrices $Q$.
For generic value of $q$ on the unit circle, (or equivalently for a generic 
 value of the level $k$) the solution of \eref{eq:3.19} is unique and 
corresponds to a diagonal pairing of the two sectors (a diagonal modular 
invariant). For special values 
of the level $k$ there exist also other solutions which correspond to 
non-diagonal modular invariants. Let us first consider the generic diagonal case.
 The solution of the braid invariance condition  \eref{eq:3.19} 
 is   \cite{[63]}
\begin{eqnarray}
 Q_{\mu\nu}(I_1,I_2,I_3,I_4) \enspace &=& 
 (-1)^{\mu+\nu}{[\mu]![\nu]![\mu-I_{12}+I_{34}]![\nu-I_{12}+I_{34}]!
\over
[2I_1]![2I_2]![2I_3]![2I_4]!} \nonumber \\
\ms
& \times & \sum_{\rho=0}^{\min\left(\mu,\nu,k(I)\right)}
T_\rho\left(\mu,\nu;I_i\right)\ , 
\label{eq:3.64a}
\end{eqnarray}
where  $\mu,\nu=0,\ldots,2I_4$,
$$
k(I) = k-I_1-I_2-I_3+I_4  
$$
and \begin{eqnarray}
 T_\rho(\mu,\nu;I_i) &=& [2I_{34}+2\rho+1] \nonumber \\
 \ms
& \times & {[I_1+I_2+I_{34}+\rho+1]! [I_1+I_2-I_{34}-\rho]![2I_4-\rho]!
 \over
[2I_{34}+\mu+\rho+1]![2I_{34}+\nu+\rho+1]! }\  \nonumber \\ 
\ms
& \times & { 
[2I_{34}+\rho]![I_{12}+I_{34}+\rho]![2I_3+\rho+1]!\over
 [\mu-\rho]![\nu-
\rho]![\rho]![I_{34}-I_{12}+\rho]!}\ . 
\nonumber
\end{eqnarray}
It is straightforward but rather lengthy to check that  \eref{eq:3.64a} satisfies  \eref{eq:3.19}.

In order to study the factorization properties 
of the two dimensional Green functions 
let us rewrite the 4-point chiral conformal blocks 
in the tree bases. In the s-channel, which  exhibits the singularities of the 
solutions for small $z_{12}$ (hence small  $\eta$) we find  
\begin{eqnarray}
S_{I_{34}+\lambda}^{(I_1,I_2,I_3,I_4)}(z,\zeta) &=&
\sum_{\nu=0}^{2I_4}
w_\nu^{(I_1,I_2,I_3,I_4)} (z,\zeta) \ 
\sigma_{\nu\lambda}^{-1}(I_1,I_2,I_3,I_4) \nonumber \\
\ms
& = &
\sum_{\nu=\lambda}^{2I_4} { (-1)^{\nu-\lambda} [\nu]! [\nu-I_{12}+I_{34}]!
[2I_{34}+2 \lambda +1]! \over [\nu-\lambda]![\lambda]!
[\lambda-I_{12}+I_{34}]! [2I_{34}+\nu+\lambda+1]!} \nonumber \\
\ms
&& \times w_\nu^{(I_1,I_2,I_3,I_4)} (z,\zeta) \ .
\label{eq:3.67}
\end{eqnarray}
Let us stress that for $q$ a root of unity (note that $q^{k+2}=-1$) the matrix elements 
of the matrix $\sigma^{-1}$ are well defined only if 
\begin{equation}
I_1+I_2+I_{34}+\lambda\leq k\ . 
\label{eq:3.68}
\end{equation}
In other words the s-channel conformal blocks \eref{eq:3.67}
are well defined only for intermediate fields that respect the 
fusion rules \eref{eq:2.56a}. 
In the rest of the paper, we shall use  \eref{eq:3.67}
and all other tree bases formulae only for such
intermediate fields. Having this in mind we can 
introduce also the matrix formally inverse to 
$\sigma^{-1}$
\begin{equation}
 \sigma_{\lambda\mu} (I_1,I_2,I_3,I_4)
={[2I_{34}+\lambda+\mu]![I_{34}-I_{12}+\lambda]!\over
[2I_{34}+2\lambda]![I_{34}-I_{12}+\mu]!}\,\left[{\lambda\atop\mu}\right] \ .
\label{eq:3.69}
\end{equation}
In the s-channel basis \eref{eq:3.67}
the exchange operator $B_1$ has a simple  diagonal form
\begin{equation}
{{((B^{s}_1)^{I_1I_2I_3I_4})}^\lambda}_\mu =
\delta^\lambda_\mu (-1)^{I_1+I_2-I_{34}-\mu}
q^{(I_{34}+\mu)(I_{34}+\mu+1)-I_1(I_1+1)-I_2(I_2+1)} \ ,
\label{eq:3.70}
\end{equation}
while the exchange operator  $B_2$ and the fusion matrix  $F$ are
given by complicated expressions. In particular for  $F$ one finds 
\begin{equation}
((F^s)^{I_1I_2I_3I_4})_{\mu \nu} = 
\sum_{\lambda=0}^{2 I} \sigma_{\nu \lambda}(I_1,I_2,I_3,I_4)
\sigma^{-1}_{2I - \lambda, \mu}(I_3,I_2,I_1,I_4) \ , 
\label{eq:3.71a}
\end{equation}
where  $I=I_4$ (see equation \eref{eq:3.30}).
Inserting the expressions for $\sigma$ and $\sigma^{-1}$ we obtain
\begin{eqnarray}
&& ((F^s)^{I_1I_2I_3I_4})_{\mu \nu}  = \!\!\!\!\!
\sum_{\lambda=0}^{{\rm min}(\nu, 2I -\mu )} \!\!\!\!\!
(-1)^{2I-\lambda-\mu}
{[2I_{34}+\nu+\lambda]![I_{34}-I_{12}+\nu]![\nu]!
\over
[2I_{34}+2\nu]![I_{34}-I_{12}+\lambda]![\nu-\lambda]![\lambda]!} 
\nonumber \\
\bs
&& \quad \times
{[2I-\lambda]![2I-\lambda-I_{32}+I_{14}]![2I_{14}+2\mu+1]! 
\over 
[2I-\lambda-\mu]![\mu]![\mu-I_{32}+I_{14}]![2I_{14}+2I-\lambda+\mu+1]!} \ .
\label{eq:3.71b}
\end{eqnarray}

The other tree basis, the u-channel,  exhibits the singularities of the 
solutions for small $z_{23}$ (hence small  $1-\eta$). To construct it let 
us note that the  
KZ equation  as differential equation in  $1-\eta$
for the conformal blocks with isospin order $I_3 I_2 I_1 I_4$
coincides with the KZ equation  in  $\eta$
for the conformal blocks with isospin order $I_1 I_2 I_3 I_4$. 
Thus we can  define the u-channel blocks which diagonalize 
the exchange operator $B_2$ as 
\begin{eqnarray}
&& U_{I_{14}+\lambda}^{(I_1,I_2,I_3,I_4)}(\eta) =
S_{I_{14}+\lambda}^{(I_2,I_3,I_4,I_1)}(1-\eta)  =  
\
(-1)^{I_2+I_3-I_1-I_4} \nonumber \\
\ms
&& \times q^{I_1(I_1+1)+I_4(I_4+1)-I_2(I_2+1)-I_3(I_3+1)}
S_{I_{14}+\lambda}^{(I_3,I_2,I_1,I_4)}(1-\eta) \ ,
\label{eq:3.72}
\end{eqnarray}
where the second equation follows from  the diagonal form of the exchange operator 
$B_1$ \eref{eq:3.70}, (and hence also of  $B_3$ and  
$B_1 B_3^{-1}$) in the s-channel basis.
The u-channel blocks \eref{eq:3.72} are related to the 
s-channel blocks \eref{eq:3.67} by the fusion matrix $F^s$ \eref{eq:3.71b}
\begin{equation}
U_{I_{14}+\mu}^{(I_1,I_2,I_3,I_4)}(\eta) = 
\sum_{\nu} ((F^s)^{I_1I_2I_3I_4})_{\mu \nu} 
S_{I_{34}+\nu}^{(I_1,I_2,I_3,I_4)}(\eta) \ .
\label{eq:3.73}
\end{equation}

The two dimensional Green functions can be expressed 
in terms of the tree conformal blocks  as 
\begin{eqnarray}
G_4^{(I_i)}(z,\overline{z};\zeta,\overline{\zeta}) &=&
\sum_{\nu=0}^{\min(2I,k(I))} [g_s]_\nu^{(I_i)}
\overline{S}_{I_{34}+\nu}(\overline{z},\overline{\zeta})
S_{I_{34}+\nu}(z,\zeta)\  \nonumber \\
\bs
&=&
\sum_{\mu=0}^{\min(2I,k(I))} [g_u]_\mu^{(I_i)}
\overline{U}_{I_{14}+\mu}(\overline{z},\overline{\zeta})
U_{I_{14}+\mu}(z,\zeta) \  
\label{eq:3.74}
\end{eqnarray}
The normalization constants  $[g_s]$ and $[g_u]$ can be written as 
\begin{eqnarray}
{[ g_s ]}_\nu^{(I_1I_2I_3I_4)} &=&  {C_{I_1 I_2 I_{34}+\nu} 
C_{I_3 I_4 I_{34}+\nu} \over N_{I_{34}+\nu}} \  \label{eq:3.75a} \\
\bs
 {[ g_u ]}_\mu^{(I_1I_2I_3I_4)} &=& {C_{I_2 I_3 I_{14}+\mu} 
 C_{I_1 I_4 I_{14}+\mu} \over N_{I_{14}+\mu}} \ 
\label{eq:3.75b}
\end{eqnarray}
where  $C_{I_1 I_2 I_3}$ are 
\begin{eqnarray}
 C_{I_1 I_2 I_3} &=&  [I_1+I_2+I_3+1]!
\nonumber \\
\ms
&& \times {[I_1+I_2-I_3]![I_2+I_3-I_1]![I_1+I_3-
I_2]!\over[2I_1]![2I_2]![2I_3]!} \ , 
\label{eq:3.76}
\end{eqnarray}
while  $N_I$ is equal to  
\begin{equation}
N_I = C_{I I 0} = [2I+1] \ .
\label{eq:3.77}
\end{equation}

Now we can impose the factorization property 
in both the s- and the u- channels.
Comparison of \eref{eq:3.75a} and \eref{eq:3.75b} 
with the two dimensional OPE \eref{eq:3.21blabla}
shows that the two dimensional structure constants
in the diagonal model (in which the two dimensional fields have equal chiral and antichiral 
labels) are given by \index{structure constants ! bulk}
\begin{equation}
C_{(I  I) (J  J)}^{(K  K)} =  
{C_{I J K } \over N_K} \ .
\label{eq:3.79}
\end{equation}
Moreover, if we choose the normalizations of the 2-point functions to be equal to  $N_I$
\eref{eq:3.77},  the normalizations of the 3-point functions are equal to   
$C_{I_1 I_2 I_3}$ \eref{eq:3.76}.

This construction can be extended also to the 
non-diagonal $SU(2)$ current algebra models.
For the $D_{odd}$ series of models, which exist  for 
values of the level $k=4 p -2$, the structure constants are  
\index{structure constants ! bulk}
\begin{equation}
C_{(I \bar I) (J \bar J)}^{(K \bar K)} =  \epsilon_{(I \bar I) (J \bar
J) (K \bar K)}
\sqrt{C_{I J K } C_{\bar I \bar J \bar K } \over N_K N_{\bar K}} \ ,
\label{eq:3.80}
\end{equation}
where the  signs  $\epsilon$ are symmetric in all three pairs of indices and 
differ from $+1$ only if two pairs of the isospins, say  $I, \bar I , J , \bar J $,  
are half integers, in which case they are equal to    $(-1)^K \ ( = (-1)^{\bar K})$.

For the other $SU(2)$ current algebra models denoted by $D_{even}$ and  $E_6$, 
$E_7$, $E_8$
one can also compute the structure constants \cite{[90]}. The resulting expressions 
are not as simply related to the diagonal ones. 
This can be explained by the fact that  these models correspond to 
extensions of the observable algebra, so their structure is determined
by this extension, rather than by the underlying  $SU(2)$ current algebra.


\section{CFT on surfaces with holes and crosscaps}
\label{sec4}

Conformal field theories 
in presence of boundaries 
have been introduced by Cardy  to describe critical 
phenomena in Statistical Mechanics and solid state physics
\cite{[93],[94],[20]}.
An alternative  approach, called 
open and unoriented descendants  construction, was  proposed
by Sagnotti 
 in the framework of string theory  
to unify in a consistent way  open strings with closed 
oriented and unoriented strings \cite{[92]}.
In this section we shall review some general properties 
of boundary CFT. The $SU(2)$ conformal current algebra
models will be again used as an example. On one hand, they are 
relatively simple and all the necessary data (chiral conformal
blocks, structure constants, exchange operators) are explicitly known.
On the other hand the $SU(2)$ models exhibit many 
features of the general case (like infinite series of nondiagonal
models and non abelian fusion rules).

\subsection{Open sector, sewing constraints}

The presence of a boundary breaks the two dimensional conformal symmetry, since
the boundary cannot be invariant under all the transformations of 
 $Vir \otimes {\overline {Vir}}$.  
If the central charges of the two chiral algebras 
are equal $\bar c = c$, it is possible to introduce boundaries which are preserved 
at most by the diagonal subalgebra $Vir_{diag}$. 
We shall call such boundaries conformal boundaries.
In the rest we shall assume that all boundaries are conformal.
The introduction of non conformal boundaries is also possible, but one cannot use anymore
the tools of conformal field theory for their study.  

Assume that the conformal boundary coincides 
with the line $x^1 = 0 $. 
The conformal invariance condition means that there is no energy transfer 
across the boundary, hence the stress energy tensor satisfies 
 \cite{[93]}
\begin{equation}
\Theta(x_-) = \bar \Theta (x_+) \quad {\rm for} \quad x_- = x_+  \quad 
\Leftrightarrow  \ x^1 = 0 \ ,
\label{eq:4.1} 
\end{equation}
 since $x_{\pm}=x^0 \pm x^1$.
So one can define the stress energy tensor in the theory with 
conformal boundaries as

\begin{equation}
\Theta _d(x) = \left\{ {\Theta (x_-) \ {\rm for} \ x^1 \geq 0 }
\atop {\bar \Theta (x_+) \ {\rm for} \ x^1 <0 }
\right. \  
\label{eq:4.2}
\end{equation}

In a similar way, if the two dimensional theory is invariant under the product  
of two isomorphic conformal current algebras ${\cal A}\otimes\bar{\cal A}$
with equal levels $\bar k = k $, the boundary can be preserved at most  
by the diagonal subalgebra ${\cal A}_{diag}$. 
Such boundaries are called symmetry preserving, 
the currents in this case 
are   defined as 
\begin{equation}
j^a_d(x) = \left\{ {j^a(x_-) \ {\rm for} \ x^1 \geq 0 }
\atop {\bar j^a( x_+) \ {\rm for} \ x^1 <0 }
\right. \  
\label{eq:4.3}
\end{equation}
One can introduce also conformal boundaries that are preserved only by 
a proper subalgebra  ${\cal A}^{\prime} \subset {\cal A}_{diag}$ (such that the boundary is still 
invariant under $Vir_{diag}$).
Such boundaries are called symmetry breaking (or symmetry non preserving)
boundaries and have been also studied  \cite{[Brokenbulk]}.
In these lectures we shall restrict our attention only to  the simpler case 
of symmetry preserving boundaries.

We can pass to the analytic picture by mapping the boundary onto the
unit circle by a Cayley transform \eref{eq:1.15}. The stress energy tensor becomes
\begin{equation}
T_d(z) = \left\{ {T(z) \ {\rm for} \ \vert z \vert \leq 1 }
\atop { {1 \over { z^4} } \bar T({1 \over z}) \ {\rm for} \ \vert z
\vert >1  }
\right. \  
\label{eq:4.4}
\end{equation}
(where we used    $ \bar z \leftrightarrow 1/z$ in this picture),
while the currents are 
\begin{equation}
J^a_d(z) = \left\{ {J^a(z) \ {\rm for} \ \vert z \vert \leq 1 }
\atop {-{1 \over { z^2} } \bar  J^a({1 \over z}) \ {\rm for} \ \vert z
\vert >1  }
\right. \   
\label{eq:4.5}
\end{equation}
The sign change with respect to  \eref{eq:4.4} 
comes from the prefactor in the Cayley transform.

Let us introduce also the following combinations of the 
Laurent modes of the stress energy tensor $T$ and the currents $J^a$
\begin{equation}
{\cal L }_n = L_n - \bar L_{-n}  \ 
\label{eq:4.6}
\end{equation}
and 
\begin{equation}
{{\cal J}^a }_n = J^a_n + \bar J^a_{-n} \ .
\label{eq:4.7}
\end{equation}
Since the left and right central charges and  levels are equal 
($\bar c = c$, $\bar k = k$)
the modes  \eref{eq:4.6} satisfy the commutation relations of the Virasoro algebra 
with  central charge equal to zero 
\begin{equation}
\left[ {\cal L }_n , {\cal L }_m \right] = 
(n-m) {\cal L }_{n+m} \ ,
\label{eq:4.8}
\end{equation}
while the modes \eref{eq:4.7} 
satisfy  the commutation relations of the current algebra with  level equal to zero 
\begin{equation}
\left[ {{\cal J}^a }_n , {{\cal J}^b }_m \right] = \rmi f^{abc} {{\cal J}^c
}_{n+m} \ . 
\label{eq:4.9}
\end{equation}
These two algebras have no nontrivial representations, hence the modes  
 \eref{eq:4.6}, \eref{eq:4.7} annihilate all the boundary states
\index{state ! boundary}
  $\vert B \rangle$ in the theory
\begin{equation}
{\cal L }_n\vert B \rangle \ = \ (L_n - \bar L_{-n} )\vert B \rangle \ = \ 0 \ 
\label{eq:4.10}
\end{equation}
and
\begin{equation}
{{\cal J}^a }_n\vert B \rangle \ = \ (J^a_n + \bar J^a_{-n} )\vert B \rangle \ = \ 0 \ .
\label{eq:4.11}
\end{equation}
For rational models a basis of states that satisfy  \eref{eq:4.10} called Ishibashi states
\index{state ! Ishibashi}
has been constructed in  \cite{[96]} 
as infinite sums of products of left and right states
\begin{equation}
\vert {\cal I}_{\Lambda} \rangle \ = \ \sum_m \vert \Lambda, m \rangle \otimes
\overline{\vert \Lambda, m \rangle} \ , 
\label{eq:4.12}
\end{equation}
where the sum extends over all the quasiprimary  descendants of the 
primary state  $\vert \Lambda \rangle$.
Note that the Ishibashi states are not eigenvalues of the energy 
$L_0 + \bar L_0$ and are not normalizable in the usual sense.

One important consequence of equations \eref{eq:4.4}, \eref{eq:4.5}
is that in the presence of boundaries 
the $n$-point functions of two dimensional primary  fields 
$\phi_{\Lambda\bar\Lambda}(z,\bar z)$ 
and the  chiral conformal blocks of $2n$-chiral vertex operators with the same weights
satisfy the same equations as functions of the  $2n$ variables $(z_1,\bar z_1, \dots , z_n, \bar z_n)$
\cite{[93]}. 
Indeed, since the chiral and antichiral parts of the
stress energy tensor and of the currents act independently on the chiral and antichiral 
vertex operators in the decomposition of the two dimensional primary fields
  (see also equation  \eref{eq:2.45})
\begin{equation}
\phi_{\Lambda\bar\Lambda}(z,\bar z) =\sum_{{\Lambda_i\ \ \bar\Lambda_i\atop
\Lambda_f\ \ \bar\Lambda_f}}\ V\kern-5pt\textstyle{{\Lambda_f\atop\Lambda\ \
\Lambda_i}}(z)\
\bar V\kern-5pt\textstyle{{\bar\Lambda_f\atop\bar\Lambda\ \ \bar\Lambda_i}}(\bar
z) \ 
 n_{i \bar i }^{f \bar f } \ ,
\label{eq:4.13}
\end{equation}
 the $n$-point functions of the two dimensional fields in the theory 
with boundaries are linear combinations of $2n$-point chiral conformal blocks.
Note the difference with respect to the case without boundaries
reviewed in the previous section, where the 
 two dimensional functions are sesquilinear combinations of $n$-point
conformal blocks.

Another important property of the boundary is the existence of one
dimensional fields
 $\psi$  called boundary fields \cite{[93]}.
\index{field ! boundary}
They are defined only on the boundary (on the unit circle in the analytic picture). 
Equations \eref{eq:4.4}, \eref{eq:4.5} imply that the Virasoro algebra 
and the conformal current algebra which act on the boundary have the same
central charge and the same level as the chiral ones. 
Hence the primary boundary fields can be labelled  by the same set of weights $\Lambda$.
There can be different boundary conditions on different portions of the boundary,
which we  denote by labels $a,b,c$. The boundary fields 
carry two boundary condition labels $\psi^{ab}_{\Lambda}(x)$  and 
change the boundary condition from $b$ to $a$. 
In general also a degeneracy label accounting for the multiplicity
of the boundary fields may be necessary. For simplicity we shall omit the 
degeneracy labels. For a more accurate analysis of this point see e.g.\cite{[PZNP]}.
We shall 
denote the argument of the boundary fields by   
$x$ which takes values only on the unit circle, 
to distinguish it from  $z$ ($\bar z$) which take values inside (outside) the unit 
circle.

In general the boundary fields do not locally commute, 
rather they behave much like the 
chiral vertex operators under the exchange algebra. 
In other words in correlation functions the ordering of their arguments on the circle
cannot be changed arbitrarily. In particular, this implies that the 4-point functions of 
boundary fields satisfy only one crossing symmetry relation, called planar duality,
in contrast with the two dimensional case where crossing symmetry of the 4-point 
functions implies two duality relations.

To simplify the notation in this section we shall consider  
only the $SU(2)$ current algebra models and
label the fields by  $i=2 I_i +1$ and $\bar i = 2 \bar I_i +1$ 
rather than by their  weights $\Lambda_i$ and  $\bar \Lambda_i$.
Hence the identity operators carry the label $1$.
When this is not  ambiguous, we shall also  omit the space-time ($z$ and $x$)  
and $SU(2)$ ($\zeta$) variables.

The operator product expansion for the boundary fields schematically 
has the form (note the continuity of the boundary indices) 
\index{operator product expansion ! boundary}
\begin{equation}
{\psi_i}^{ab} \ {\psi_j}^{bc} \sim \sum_l \ C_{ijl}^{abc} \ \psi_l^{ac}  \ ,
\label{eq:4.14}
\end{equation}
where the sum is over all the values allowed by the fusion rules 
\eref{eq:2.56a}.
The boundary structure constants \index{structure constants ! boundary}
$C_{ijl}^{abc}$ are in general not symmetric. 
Other important data are the normalizations of the 2-point functions of
the boundary fields, since they cannot be chosen arbitrarily \cite{[94]}.
To define them one has to specify also the order of the arguments, since the
boundary fields do not commute.
Both variables are on the unit circle so we can 
order them by their phase 
\begin{eqnarray}
&& \langle {\psi_i}^{ab} ( x_1;\zeta_1 ) \ {\psi_i}^{ba} ( x_2;\zeta_2 ) \rangle \ = \
{{{\alpha_i}^{ab}} 
 (\zeta_{12})^{2 I_i} \over
{(x_{12})^{2\Delta_i}}} \quad  \\
\ms 
&& \qquad {\rm for} \quad Arg(x_2)  \ < \ Arg( x_1)  \ , \nonumber
\label{eq:4.15}
\end{eqnarray}
where $I_i$ is the isospin of $\psi_i$. 
The normalizations of the fields
with exchanged boundary labels are related.  
For example, 
for the $SU(2)$ current algebra models one finds   
\begin{equation}
{\alpha_i}^{ab}  \ = \ {\alpha_i}^{ba} \ (-1)^{2 I_i} \ .
\label{eq:4.16}
\end{equation}

Let us stress that even if we consider in detail only the 
$SU(2)$ conformal current algebra case, most of the formulae
are valid also in more general cases (with minor modifications in the  
numerical factors). For instance, in the unitary minimal models
case one just has to omit all the isospin dependence.

Using the boundary OPE \eref{eq:4.14} we can compute in two different ways 
the three-point functions of the boundary fields
$\langle {\psi_i}^{ab} {\psi_j}^{bc} {\psi_l}^{ca}\rangle$ and $ \langle {\psi_j}^{bc}
{\psi_l}^{ca} {\psi_i}^{ab}\rangle$. This gives the following consistency conditions  
\begin{equation}
C_{ijl}^{abc} \ {\alpha_l}^{ac} \ = \ C_{jli}^{bca} \ {\alpha_i}^{ab} \quad
{\rm and } \qquad
C_{jli}^{bca} \ {\alpha_i}^{ba} \ = \ C_{lij}^{cab} \ {\alpha_j}^{bc} \quad  ,
\label{eq:4.17a}
\end{equation}
that together with  \eref{eq:4.16} imply also 
\begin{equation}
C_{ijl}^{abc} \ {\alpha_l}^{ac} \ = \  (-1)^{2 I_i}  \
C_{lij}^{cab} \ {\alpha_j}^{bc} \quad  .
\label{eq:4.17b}
\end{equation}
The natural normalization of the boundary identity operator is 
\begin{equation}
C_{i{\bf 1}i}^{abb} \ = \ 1 \qquad   \qquad 
 \langle {\bf 1}^{aa} \rangle \ = \
\alpha_{\bf 1}^{aa}  \quad , 
\label{eq:4.18}
\end{equation}
while all other one-point functions of the boundary fields vanish.

The  planar duality constraint for the 4-point functions 
$\langle \psi_i^{ab}  \psi_j^{bc}  \psi_k^{cd}  \psi_l^{da} \rangle $ 
reads
\begin{equation}
\sum_p \ C_{ijp}^{abc} \ C_{klp}^{cda} \ \alpha_p^{ac} \ S_p(i,j,k,l) \ = \
\sum_q \ C_{jkq}^{bcd} \ C_{liq}^{dab} \ \alpha_q^{bd} \ U_q(i,j,k,l)  \ \ 
\label{eq:4.19}
\end{equation}
and after expressing the  $u$-channel blocks \eref{eq:3.72} in terms of  the 
$s$-channel blocks \eref{eq:3.67} by the fusion matrix   $F^s$ \eref{eq:3.71b} as (see also \eref{eq:3.73})
\begin{equation}
U_q (i,j,k,l) \ = \ \sum_p \ F_{qp} (i,j,k,l) \ S_p (i,j,k,l) \quad 
\label{eq:4.20}
\end{equation}
we obtain a quadratic relation for the boundary structure constants 
$C_{ijk}^{abc}$ and the 2-point normalizations
$\alpha_i^{ab}$  
\begin{equation}
C_{ijp}^{abc} \ C_{klp}^{cda} \ \alpha_p^{ac} \ = \
\sum_q \  C_{jkq}^{bcd} \ C_{liq}^{dab} \ \alpha_q^{bd} \ 
F_{qp} (i,j,k,l) \quad .
\label{eq:4.21}
\end{equation}
These relations do not determine completely
the boundary structure constants. In other words, the boundary theory 
cannot be considered independently, but only as a part of the 
two dimensional conformal theory.

The relation between the bulk and boundary fields is encoded into 
the bulk-to-boundary expansion \index{operator product expansion ! bulk-to-boundary}
\begin{equation}
  \phi_{i, {\bar i}} \Big\vert_a\ \sim \ \sum_j \ C_{(i,{\bar i})j}^a \
{\psi_j}^{aa} \quad ,
\label{eq:4.22a}
\end{equation}
that expresses the two dimensional fields in front of a portion of boundary 
with given boundary condition $a$ in terms of the corresponding boundary 
fields. The sum is again over all the values allowed by the fusion rules. 
\index{structure constants ! bulk-to-boundary}   
The proper normalization of the  identity operator gives 
\begin{equation}
 C_{({\bf 1},{\bf 1}){\bf 1}}^a \ = \ 1 \quad 
\label{eq:4.22b}
\end{equation}
for all boundary conditions $a$.

The consistency of the operator product expansions \eref{eq:3.21blabla}, \eref{eq:4.14}
 and  \eref{eq:4.22a}
have been studied by  Lewellen \cite{[97]}, who has shown that the
 complete set of relations (called also sewing constraints) 
which guarantee the consistency 
of the theory includes  two more equations, the first one involving 4-point functions and the second one
involving 5-point functions. 
The first relation arises from the correlation functions of one two
dimensional bulk field
and two boundary fields. As already stressed the boundary fields have
a fixed order of the arguments, but the two dimensional fields have to be local 
also in presence of  boundary fields, which implies 
\begin{equation}
\langle \phi_{(i,{\bar i})} \ \psi_j^{ba} \ \psi_k^{ab} \rangle \ = \ 
\langle \psi_j^{ba} \ \phi_{(i,{\bar i})} \  \psi_k^{ab} \rangle \ .
\label{eq:4.23a}
\end{equation}
Note that the bulk field is expanded in front of portions of the boundary with different boundary 
conditions in the left and  in the right hand sides of this equation. Using also 
 \eref{eq:4.14}, \eref{eq:4.22a} 
one obtains
\begin{equation}
\sum_l \
C_{(i,{\bar i}) l}^b \ C_{ljk}^{bba} \ \alpha_k^{ba} \ S_l(i,{\bar i},j,k)
\ = \ 
\sum_n \ C_{(i,{\bar i})n}^a \ C_{jnk}^{baa} \ \alpha_k^{ba} 
\ U_n(j,i,{\bar i},k)
\ .
\label{eq:4.23b}
\end{equation}
To derive the constraint on the structure constants we have to relate the 
$U$- and the $S$- blocks. A convenient way to do this is to use 
repeatedly the fusion matrix (and its inverse) in such a way that the exchange 
operators act always diagonally (see \eref{eq:3.70}). In other words before applying    
 $B_1$ or  $B_3$ we change to the s-channel basis, while before 
applying  $B_2$ we change to the u-channel basis.
The resulting composite exchange operator is  
\begin{eqnarray}
U_n (j,i,{\bar i},k) \ &=& \ \sum_{m,r,s,p,l} \ F_{nm}(j,i,{\bar i},k) \ 
( B_1 )_{mr} (i,j,{\bar i},k) \ F^{-1}_{rs}(i,j,{\bar i},k) \nonumber \\
\ms 
 & \times &( B_2 )^{-1}_{sp} ( i,{\bar i},j,k ) \ F_{pl} (i,{\bar i},j,k) \
S_l (i,{\bar i},j,k) \  . 
\label{eq:4.24}
\end{eqnarray}
Inserting this in \eref{eq:4.23b} and using the explicit expressions for 
the exchange operators $B_1$ and $B_2$ in the $SU(2)$ model, we obtain the constraint
\begin{eqnarray}
&& C_{(i,{\bar i})l}^b \ C_{j k l}^{bab} \ {\alpha_l}^{bb} \ = \
\sum_{m,n,p} \ (-1)^{(I_i - I_{\bar i} + 2 I_j + I_p - I_m )} \
{\rme}^{- \rmi \pi ( \Delta_i - \Delta_{\bar i} - \Delta_m + \Delta_p )} 
 \nonumber \\
 \ms
  \times && \quad C_{(i,{\bar i})n}^a \ C_{kjn}^{aba} \ {\alpha_n}^{aa}
 {F_{nm}}(j,i,{\bar i},k) \ {F^{-1}}_{mp}(i,j,{\bar i},k) \
F_{pl}(i,{\bar i},j,k)  . 
\label{eq:4.25}
\end{eqnarray}
The other independent relation can be derived from the 5-point functions
of two bulk fields and one boundary field of the form 
\begin{equation}
\langle \phi_{(i,{\bar i})} \
\phi_{(j,{\bar j})} \ {\psi_k}^{aa} \rangle \  . 
\label{eq:4.26a}
\end{equation}
This function can again be computed in two different ways.  
We can first use the two dimensional  OPE  \eref{eq:3.21blabla}
followed by a bulk to boundary OPE  \eref{eq:4.22a}, alternatively 
we can use twice the  bulk to boundary OPE  \eref{eq:4.22a} followed by 
a boundary OPE \eref{eq:4.14}.
Before proceeding we have to define a basis in the space of 5-point functions.
We shall use a tree representation which decomposes the 5-point functions 
into products of a 4-point function and a 3-point function (denoted by $g(1,2,3)$) with 
one common external leg  
\begin{eqnarray}
X_{pq}(1,2,3,4,5) \ & = &\ S_p(1,2,q,5)  \ g(q,3,4) \nonumber \\
\ms
& = & \ g(1,2,p) \ U_q(p,3,4,5) \quad . 
\label{eq:4.26b}
\end{eqnarray}
In this notation the equivalence of the two ways of computing the function 
 \eref{eq:4.26a} implies
\begin{eqnarray}
&& \sum_{p,q} \ C_{(i,{\bar i})(j,{\bar j})}^{ (p,{\bar q})} \
C^a_{(p,{\bar q})k} \ \alpha_k^{aa} \ X_{p{\bar q}}(j,i,{\bar i},{\bar j},k)
\nonumber \\ 
\ms
= \ &&  \sum_{p,q} \ C^a_{(i,{\bar i})p} \ C^a_{(j,{\bar j})q} \ 
C_{pqk}^{aaa} \ \alpha_k^{aa} \ X_{pq}(i,{\bar i},j,{\bar j},k)
\quad .
\label{eq:4.26c}
\end{eqnarray}
The two expressions can again be related by the exchange operators,
for example by  
$F_{[2]} B_1^{-1} B_2B_1B_2 F_{[2]}^{-1}$, 
where the label in brackets indicates on which 
4-point subtree  acts  the fusion matrix  $F$. 
We can use the 
Yang-Baxter equations \eref{eq:3.53}  to express  $B_2B_1B_2$ in terms of $F_{[1]}$
obtaining (if $\alpha_k^{aa} \neq 0 $)
\begin{eqnarray}
&& C_{(i,{\bar i})(j,{\bar j})}^{ (p,{\bar q})} \ C^a_{(p,{\bar q})k} \ = \
\sum_{r,s,t} \ (-1)^{( I_j - I_t + I_r )} \ 
{\rme}^{- \rmi \pi ( \Delta_j - \Delta_t + \Delta_r )}  \ C_{rsk}^{aaa} \nonumber \\
\ms
&& \times    C^a_{(i,{\bar i}) r} \ C^a_{(j,{\bar j})s} \  
 F_{st}(r,j,{\bar j},k) \ F_{rp}(j,i,{\bar i},t) \ 
F_{t{\bar q}}^{-1}(p,{\bar i},{\bar j},k) .  
\label{eq:4.27}
\end{eqnarray}

Both equations \eref{eq:4.25} and \eref{eq:4.27} can be written in several 
different equivalent forms, since the exchange operators satisfy duality relations 
(like the Yang--Baxter equation \eref{eq:3.53}) \cite{[61]}.
Our derivation follows \cite{[98]}. For alternative ones see also \cite{[97],[PZNP],[Runkel]}. 
In particular  \cite{[Runkel]} contains 
  the general solution of the sewing constraints  
for the unitary minimal models with a detailed analysis of  the residual 
normalization freedom.
Here we shall address only a simpler problem, namely 
we shall try to count the allowed 
boundary conditions. Note that
 in all the sewing constraints the 
boundary fields enter as external insertions, so one can always 
start with only one type of boundary labels, say $a$, and solve 
only the corresponding subsystem.  
There is however  a systematic way to determine also the 
whole set of allowed  boundary conditions. In other words, only by analyzing the 
sewing constraints one can find all boundary states $\vert a \rangle$.
To illustrate this point, let us consider one particular case of the function 
\eref{eq:4.26a}, namely 
$ \langle \phi_{(i,{\bar i})} \ \phi_{(j,{\bar j})} \ {\bf 1}^{aa}  \rangle$.
Then the condition 
\eref{eq:4.27} becomes 
\begin{eqnarray}
C_{(i,{\bar i})(j,{\bar j})}^{ (q,{\bar q})} \ C^a_{(q,{\bar q}){\bf 1}} 
\ \alpha_{\bf 1}^{aa} \ &=& \
\sum_{p} \ (-1)^{( I_j - I_{\bar j} + I_p )} \ 
{\rme}^{- \rmi \pi ( \Delta_j - \Delta_{\bar j} + \Delta_p )}  \nonumber \\
\ms
& \times & \alpha_p^{aa} \
C^a_{(i,{\bar i})p} \ C^a_{(j,{\bar j})p} \
F_{pq}(j,i,{\bar i},{\bar j})   \quad . 
\label{eq:4.28}
\end{eqnarray}
Multiplying by $F^{-1}_{qr}(j,i,{\bar i},{\bar j})$, summing on $q$ 
and keeping only the equation corresponding to $r=1$ we find a system of 
equations for the bulk-to-boundary coefficients in front of  the 
boundary identity  operators
\begin{equation}
{B}_{i}^a = C^a_{(i,{\bar i}){\bf 1}}  \ , 
\label{eq:4.29}
\end{equation}
where to simplify notation we used that for a permutation 
modular invariant the antichiral label of a field $\bar i$ is 
determined by its chiral label $i$. 
The resulting relation has the form 
\begin{equation}
{B}_i^a \ {B}_j^a \ = \ \sum_l \ 
 \ {X_{ij}}^l \ {B}_l^a
\quad ,
\label{eq:4.29b}
\end{equation}
for all $a$ with $a$-independent structure constants ${X_{ij}}^l$ 
that vanish if the fusion rules ${N_{ij}}^l$ are zero.
The number of different solutions of these equations 
determines also the number of allowed boundary conditions.
In order to compute the values of the structure constants 
${X_{ij}}^l$ one needs to know the two dimensional structure constants 
and the expressions for the fusion matrix in the model. 
As already stressed, these data are known only in a very restricted 
number of cases. In order to bypass this difficulty, in \cite{[Classifying]}
an alternative approach
was proposed. One can postulate that 
\eref{eq:4.29b} holds and that the structure constants ${X_{ij}}^l$
form a commutative and associative algebra, called Classifying algebra. 
\index{algebra ! classifying} Then  the 
reflection coefficients ${B}_i^a$ are given by the  
representations of this algebra which in some cases can be explicitly found.

For the  $SU(2)$ case from the 
explicit expressions of the fusion matrix 
 \eref{eq:3.71b} and the two dimensional structure constants  \eref{eq:3.80}, 
we can compute the values of ${X_{ij}}^l$ both in the diagonal $A$ 
models and in the non-diagonal $D_{odd}$ models, obtaining  
\begin{equation}
{B}_i^a \ {B}_j^a \ = \ \sum_l \ 
\epsilon_{ijl} \ {N_{ij}}^l \ {B}_l^a
\quad ,
\label{eq:4.30}
\end{equation}
where the signs  $\epsilon_{ijl}$, 
 present only for the $D_{odd}$ models, 
are defined after equation \eref{eq:3.80} 
(they are symmetric in all three indices and 
are equal to $(-1)$ only if two of the isospins are half integer, while the third isospin 
is an odd integer).

As an illustration we shall write down the solutions of the system \eref{eq:4.30}
in the two $SU(2)$ models of level $k=6$. In the diagonal $A$ 
model there are seven different solutions 
for the reflection coefficients ${B}_i$,
which are reported in   \tref{tt1}.
\begin{table}
\caption{Reflection coefficients for the diagonal $SU(2)$ level $k=6$ model}
\label{tt1}
\begin{center}
\begin{tabular}{@{}llllllll@{}}
\br
a & $B_1$ & $B_3$ & $B_5$ & $B_7$ & $B_2$ & $B_4$ & $B_6$  \\
\mr
1 & 1 & $1+\sqrt{2}$ & $1+\sqrt{2}$ & 1 & $\sqrt{2+\sqrt{2}}$ & $\sqrt{2(2+\sqrt{2})}$ & $\sqrt{2+\sqrt{2}}$  \\
2 & 1 & 1 & -1 & -1 & $\sqrt{2}$ & 0  & $-\sqrt{2}$  \\
3 & 1 & $1-\sqrt{2}$ & $1-\sqrt{2}$ & 1 & $\sqrt{2-\sqrt{2}}$ & $-\sqrt{2(2-\sqrt{2})}$  & $\sqrt{2-\sqrt{2}}$  \\
4 & 1 & -1 & 1 & -1 & 0 & 0 & 0  \\
5 & 1 & $1-\sqrt{2}$ & $1-\sqrt{2}$ & 1 & $-\sqrt{2-\sqrt{2}}$ & $\sqrt{2(2-\sqrt{2})}$  & $-\sqrt{2-\sqrt{2}}$  \\
6 & 1 & 1 & -1 & -1 & $-\sqrt{2}$ & 0 & $\sqrt{2}$  \\
7 & 1 & $1+\sqrt{2}$ & $1+\sqrt{2}$ & 1 & $-\sqrt{2+\sqrt{2}}$ & $-\sqrt{2(2+\sqrt{2})}$  & $-\sqrt{2+\sqrt{2}}$  \\
\br
\end{tabular}
\end{center}
\end{table}
Note that in the diagonal models the number of boundary conditions is always 
equal to the number of two dimensional fields.

In the non-diagonal $D_5$ model,  
with torus partition function 
\begin{equation}
Z_T^{D_5} = |\chi_1|^2 + |\chi_3|^2 + |\chi_5|^2 + |\chi_7|^2 + |\chi_4|^2
+\chi_2 {\bar \chi_6} + \chi_6 {\bar \chi_2} \quad 
\label{eq:4.53}
\end{equation}
two of the coefficients (${B}_{2}$ and ${B}_{6}$) 
vanish, since the corresponding two dimensional fields are non-diagonal, 
while the presence of the signs $\epsilon_{ijl}$ modifies the equations for 
${B}_{4}$ as follows
\begin{eqnarray}
&&{B}_{4} \ {B}_{2I+1} \ = \ (-1)^I {B}_{4} \quad ,\nonumber \\
\ms
&&{B}_{4} \ {B}_{4} \ = \ {B}_{1} - {B}_{3} + 
{B}_{5} - {B}_{7} \quad .
\label{eq:4.31}
\end{eqnarray}
Hence there are only five different solutions for the reflection coefficients
${B}_i$, which are reported in  \tref{tt2}. 
\begin{table}
\caption{Reflection coefficients for the non-diagonal $SU(2)$ level $k=6$ model}
\label{tt2}
\begin{center}
\begin{tabular}{@{}llllll@{}}
\br
a & $B_1$ & $B_3$ & $B_5$ & $B_7$  & $B_4$   \\
\mr
1 & 1 & 1 & -1 & -1  & 0    \\
2 & 1 & $1+\sqrt{2}$ & $1+\sqrt{2}$ & 1  & 0   \\
3 & 1 & -1 & 1 & -1  & 2   \\
4 & 1 & -1 & 1 & -1  & -2   \\
5 & 1 & $1-\sqrt{2}$ & $1-\sqrt{2}$ & 1  & 0    \\
\br
\end{tabular}
\end{center}
\end{table}

\noindent
Note that again the number 
of different boundary conditions is equal to the number of 
two dimensional fields with charge conjugate chiral and  antichiral labels  
(or equivalently to the number of different 
Ishibashi states \eref{eq:4.12}). This
in fact is a general property
of the two dimensional conformal theories with boundaries \cite{[Classifying],[PZ PL]}.

\subsection{Closed unoriented sector, crosscap constraint}

To study the behaviour of the two dimensional fields on non-oriented
surfaces let us first introduce the crosscap. The crosscap 
is the projective plane and can be represented as a unit disc with diametrically
opposite points identified. Two dimensional surfaces  with crosscaps cannot be 
oriented. For example the Klein bottle is topologically equivalent to 
a cylinder terminating at two crosscaps.

Our analysis will follow  closely the one in the boundary case.
Like the boundaries, the crosscap breaks the two dimensional conformal symmetry
since it is not invariant under all transformations of  $Vir \otimes {\overline {Vir}}$.
If the central charges of the two algebras are equal ($\bar c = c$) there exist
 crosscaps that are preserved at most by the diagonal subalgebra $Vir_{diag}$.
Let us  again pass to the analytic picture mapping the boundary of the
crosscap onto the unit circle. Then the crosscap implies the identification 
$ \bar z  \leftrightarrow -1/z$. 
Similarly to the case of a boundary, the absence of energy flux through the crosscap allows 
to define the stress energy tensor as 
\begin{equation} 
T_d(z) = \left\{ {T(z) \ {\rm for} \ \vert z \vert \leq 1 }
\atop { {1 \over { z^4} } \bar T(-{1 \over z}) \ {\rm for} \ \vert z
\vert >1  }
\right. \  
\label{eq:4.34}
\end{equation}
while  the currents are 
\begin{equation} 
J^a_d(z) = \left\{ {J^a(z) \ {\rm for} \ \vert z \vert \leq 1 }
\atop {-{1 \over { z^2} } \bar  J^a(- {1 \over z}) \ {\rm for} \ \vert z
\vert >1  }
\right. \  
\label{eq:4.35}
\end{equation}
The combinations of the Laurent modes of the stress energy tensor and of the currents
that satisfy the Virasoro algebra with vanishing central charge  \eref{eq:4.8}
and the current algebra of zero level \eref{eq:4.9} are in this case
\begin{equation} 
{\cal L }_n = L_n - (-1)^n \bar L_{-n}  \ 
\label{eq:4.36}
\end{equation}
and
\begin{equation} 
{{\cal J}^a }_n = J^a_n + (-1)^n \bar J^a_{-n} \ .
\label{eq:4.37}
\end{equation}

The crosscap states
\index{state ! crosscap} $\vert C \rangle$ \cite{[99]} in the theory are  annihilated 
by the modes \eref{eq:4.36} and \eref{eq:4.37}
\begin{equation} 
{\cal L }_n\vert C \rangle \ = \ (L_n - (-1)^n \bar L_{-n} )\vert C \rangle \ = \ 0 \ 
\label{eq:4.38}
\end{equation}
and
\begin{equation} 
{{\cal J}^a }_n\vert C \rangle \ = \ (J^a_n + (-1)^n \bar J^a_{-n} )\vert C \rangle \ = \ 0
\ .
\label{eq:4.39}
\end{equation}
These equation have the same number of solutions as the corresponding equations  
\eref{eq:4.10} and \eref{eq:4.11} for the boundary states and one can explicitly 
construct the Ishibashi-type crosscap states like in \eref{eq:4.12}.
There is however an important difference with the boundary case, since the 
consistency conditions imply the crosscap constraint \cite{[100],[102]}, which singles out 
one  crosscap state $\vert C \rangle$. Let us stress that in general 
there may be several different 
crosscap states corresponding to different actions of the involution
$ \Omega \ : \  z  \leftrightarrow -1/z$ on the fields. The crosscap
constraint tells us only  that two different crosscap states
cannot exist simultaneously in the same theory.  

Just like in the boundary case,  the presence of a crosscap implies that 
 the $n$-point functions of the two 
dimensional primary fields are linear combinations of the $2n$-point
chiral conformal blocks.
However, in contrast with  the boundary case one cannot introduce non-trivial 
crosscap operators, since the involution $ \Omega \ : \  z  \leftrightarrow -1/z$
has no fixed points. 
In particular  only the identity operator
(which has no $z$ dependence and hence is the only   invariant under $\Omega$ one)
can contribute to the expansion of a two dimensional
primary field in front of a crosscap 
\begin{equation} 
\phi_{\Lambda \bar \Lambda}( z, \bar z) \Big\vert_{crosscap} \ \sim \ 
\Gamma_{\Lambda \bar \Lambda}  
\delta_{\Lambda {\bar \Lambda}^{C}} \ 1 \quad . 
\label{eq:4.40}
\end{equation}
Here $\Gamma_{\Lambda \bar \Lambda}$  is a normalization constant and 
${\bar\Lambda}^{C}$ is the charge conjugate of  $\bar \Lambda$. 
Let us stress that the expansion \eref{eq:4.40} can be used only for the 
computation of the one point functions of the fields in front of a crosscap.
The reason is that the operator product expansions are valid only if the arguments 
 can be connected without encountering other singularities, but in all 
$n \geq 2$ point functions in front of a crosscap  $z$ and  $\bar z = -1/z$ are 
always separated by the arguments of the other fields.

The involution $\Omega$ acts on the two dimensional 
primary fields \eref{eq:4.13}  transforming the  chiral vertex operators
into antichiral ones and vice versa, and thus relating the two dimensional field 
 \eref{eq:4.13} to the field with weights and arguments exchanged
\begin{equation} 
\phi_{\bar\Lambda\Lambda}(\bar z, z) =\sum_{{\Lambda_i\ \ \bar\Lambda_i\atop
\Lambda_f\ \ \bar\Lambda_f}}\ V\kern-5pt\textstyle{{\bar\Lambda_f\atop\bar
\Lambda\ \ \bar\Lambda_i}}(\bar
z)\
\bar V\kern-5pt\textstyle{{\Lambda_f\atop\Lambda\ \
\Lambda_i}}(z) \ 
 n_{i \bar i }^{f \bar f } \ . 
\label{eq:4.41}
\end{equation}
To simplify the notation let us denote the two weights of the field by a single label
(this is unambiguous for a permutation modular invariant)
setting  $\phi_i =  \phi_{\Lambda_i\bar\Lambda_i}$
and 
$\phi_{\bar i} =  \phi_{\bar \Lambda_i \Lambda_i}$. 
The action of $\Omega$ is \cite{[101]}
\begin{equation} 
\Omega \ \phi_{i}( z, \bar z) \
 = \ \epsilon_{i} \ 
\phi_{\bar i}(\bar z, z) \ .
\label{eq:4.42a}
\end{equation}
Since  $\Omega$ is an involution, the 
$\epsilon_i$ are just signs 
\begin{equation} 
\epsilon_{i }  \ =  \ \epsilon_{\bar i } \ =  \ \pm 1 \ 
\label{eq:4.42b}
\end{equation}
which have to respect the fusion rules \eref{eq:1.59}, hence
\begin{equation} 
\epsilon_{i } \epsilon_{j } \epsilon_{k } \ = \ 1 \quad {\rm if } \quad
N_{ijk} \neq 0  \ .
\label{eq:4.42c}
\end{equation}

As an example let us again take the $SU(2)$ current algebra. In this case
the equations 
 \eref{eq:4.42c} have only two different  solutions: 
 $ \epsilon_i = +1$ for all  integer isospin fields, and 
  $\epsilon_i = \epsilon = \pm 1$ for all  half integer isospin fields  

One convenient way to compute the $n$-point functions of the 
two dimensional fields in the front of a crosscap is 
to introduce the crosscap operator 
\cite{[102]}
\begin{equation} 
{\hat C} \ = \ \sum_l \ \Gamma_l \ {| \Delta_l \rangle \langle {\bar {\Delta}_l }|
\over \sqrt{N_l}} \quad ,
\label{eq:4.43}
\end{equation}
where $N_{l}$ is the normalization constant of the two dimensional 2-point function \eref{eq:3.77}.
the operator  $\hat C$ allows to explicitly correlate the $n$-point functions 
of the two dimensional fields in presence  of a 
crosscap with the $2n$-point chiral conformal blocks 
\begin{eqnarray}
&& {\langle \phi_{1, \bar 1} .. \phi_{n, \bar n} \rangle}_C
 = 
{\langle 0\vert  \hat C \phi_{1, \bar 1} .. \phi_{n, \bar n} \vert 0 \rangle} \nonumber
 \\
 \ms
&& =     
\sum_l  {\Gamma_l \over \sqrt{N_l}} \langle 0 | V_{\Delta_1}(z_1) ..
V_{\Delta_n}(z_n) |
\Delta_l  \rangle
 \langle {\bar \Delta}_l | {\bar V}_{{\bar \Delta}_1}({\bar z}_1) ..
{\bar V}_{{\bar \Delta}_n}({\bar z}_n) | 0 \rangle \  
\label{eq:4.44}
\end{eqnarray}
The relation   \eref{eq:4.42a} for the two dimensional fields 
implies for their functions in presence  of a crosscap 
\begin{equation} 
{\langle \phi_{i, \bar i} (z_i, {\bar z}_i ) \ X \rangle}_C \ = \
\epsilon_{( i, \bar i )} \
{\langle \phi_{\bar i, i} ({\bar z}_i, z_i ) \ X \rangle}_C \quad ,
\label{eq:4.45}
\end{equation}
where  $X$ is an arbitrary polynomial in the fields.  
These equations determine the coefficients $\Gamma_n$.
In particular for the one point functions which satisfy   
\begin{eqnarray}
  { \langle \phi_{i, {\bar i}} (z, {\bar z} )  \rangle}_C \
&=& \sum_l  {\Gamma_l \over \sqrt{N_l}} \langle 0 | V_{i}(z) | \Delta_l \rangle
 \langle {\bar \Delta}_l | {\bar V}_{\bar i}({\bar z}) | 0 \rangle \  \nonumber \\
&=&   {\Gamma_{i } \over \sqrt{N_{i}}} \ \delta_{i \bar i} \ \langle 0 | V_{i}(z) \ V_{\bar i}({\bar z}) | 0
\rangle \ = \ {\langle \phi_{{\bar i}, i} ({\bar z},z )  \rangle}_C 
\label{eq:4.46}
\end{eqnarray}
equation \eref{eq:4.45} implies
the vanishing of $\Gamma_{\ell}$ for all fields on which  $\Omega$ acts nontrivially  ($\epsilon_{\ell} = -1$). 
Note that the  factor $\sqrt{N_{i}}$
in \eref{eq:4.46}  is compensated by the normalization of the chiral function
$\langle 0 | V_{i} \ V_{\bar i} | 0 \rangle$ in accord with \eref{eq:4.40}.

To derive the crosscap constraint let us apply \eref{eq:4.45} for the 
2-point functions in presence  of a crosscap. The left hand side is 
\begin{eqnarray}
 && \langle \phi_{i, \bar i} (z_1, {\bar z}_1 )  \
 \phi_{j, \bar j} (z_2, {\bar z}_2 )  \rangle_C \  \nonumber \\
 \ms
 && \quad =     \sum_l  {\Gamma_l \over \sqrt{N_l}}  \langle 0 | V_{i}(z_1) \ V_{j}(z_2)
| \Delta_l \rangle
 \langle {\bar \Delta}_l | {\bar V}_{\bar i}({\bar z}_1) \
{\bar V}_{\bar j}({\bar z}_2) | 0 \rangle \  \nonumber \\
\ms
&& \quad  =   \sum_l  \Gamma_l  \ {{\tilde C}_{( i, \bar i )( j,
\bar j )}}
^{( l, l )} \ S_{l}(z_1, z_2,{\bar z}_1,{\bar z}_2 ) \quad , 
\label{eq:4.48}
\end{eqnarray}
where  $S_l$ are the normalized $s$-channel chiral conformal blocks \eref{eq:3.67}
(note the order of the arguments  $z_i$). 
The constants  $\tilde C$  are proportional to the 
two dimensional structure constants  \eref{eq:3.80} 
\begin{equation} 
 {{\tilde C}_{( i, \bar i )( j,\bar j )}  }^{( l, l )}  \ = \ 
 {\sqrt{N_l} \  C_{( i, \bar i )(j,\bar j )}^{( l, l )}    }\quad . 
\label{eq:4.49}
\end{equation}
In the same way for the right hand side we obtain 
\begin{eqnarray}
&& \langle \phi_{\bar i  , i} ({\bar z}_1 , z_1)  \
 \phi_{j, \bar j} (z_2, {\bar z}_2 )  \rangle_C   \ 
\nonumber \\
\ms
&& \quad =
\sum_l  {\Gamma_l \over \sqrt{N_l}}  \langle 0 | V_{\bar i}({\bar z}_1) \ V_{j}(z_2)
| \Delta_l \rangle
 \langle {\bar \Delta}_l | {\bar V}_{i}( z_1) \
{\bar V}_{\bar j}({\bar z}_2) | 0 \rangle  \  \nonumber \\
\ms
&& \quad  = \sum_l  \Gamma_l  \ {\tilde C}_{( \bar i , i )( j,
\bar j)}
^{( l, l )} \ S_{l}({\bar z}_1, z_2, z_1,{\bar z}_2 ) \quad .
\label{eq:4.50}
\end{eqnarray}
The $s$-channel blocks 
$S_{l}({\bar z}_1, z_2, z_1,{\bar z}_2 )$  
are proportional to the $u$-channel blocks  
$U_{l}(z_1, z_2,{\bar z}_1,{\bar z}_2) $ (see equation \eref{eq:3.72})
and can be related to the conformal blocks 
 in \eref{eq:4.48} 
by the exchange operator $B_1 ({B_3})^{-1} F$.
Using also the explicit form of 
 $B_1$ and $B_3$ \eref{eq:3.70} we find
\begin{equation} 
S_{l}(\bar i, j, i,\bar j ) \ = \ {( - 1 )}^{{\Delta}_i - {\bar
\Delta}_i
+ {\Delta}_j - {\bar \Delta}_j} \ \sum_n \ F_{ln} ( i,j,{\bar i},{\bar j} ) \
S_{n}(i, j, \bar i,\bar j )   \quad .
\label{eq:4.51}
\end{equation}
Inserting  (\ref{eq:4.48},\ref{eq:4.50},\ref{eq:4.51}) into equation \eref{eq:4.45} 
we obtain the final form of the crosscap constraint \cite{[102]}
\begin{eqnarray} 
&&  \epsilon_{( i ,{\bar i} )} \ {( - 1 )}^{{\Delta}_i - {\bar
\Delta}_i
+ {\Delta}_j - {\bar \Delta}_j} \ \Gamma_n \ {\tilde C}_{( i, \bar i
)(j, \bar j )}
^{( n, n )}   \nonumber \\
\bs
&& \quad   = \sum_l \ \Gamma_l  \ {\tilde C}_{( \bar i , i )( j, \bar
j)}
^{( l, l )} \  F_{ln} ( i,j,{\bar i},{\bar j} ) \quad  
\label{eq:4.52}
\end{eqnarray}
for all $n$.
Applying  $\Omega$  to the second field in the two point function leads to the same equation.
Note that the crosscap constraint is linear in $\Gamma$, hence it determines 
only the ratios $\Gamma_l / \Gamma_1$. The remaining freedom is only in the 
normalization of the two dimensional identity operator in front of the crosscap
$\Gamma_1$. The simplest way to determine $\Gamma_1$ is to impose the integrality 
condition on the partition functions which we shall describe in the next section. 
An alternative approach would be to use the topological equivalence of three crosscaps 
to a handle and one  crosscap that is expected to  give a nonlinear relation for $\Gamma_l$.
The explicit form of this relation  is however still not known.


\section{Partition functions}
\label{sec5}

The two dimensional structure constants  are explicitly known only in 
a very limited number of cases. This does not allow in general to compute 
the $n$-point functions in the presence of boundaries or crosscaps and to solve
the sewing constraints.
Here we shall describe an alternative approach, proposed in \cite{[92]}
in the framework of string theory. It gives less detailed 
information about the theory, but is applicable in all cases when the modular matrices
$S$ and $T$ \eref{eq:1.54} are known. Just like modular invariant
torus partition functions are classified in 
many cases when the structure constants are not known, the 
partition function on the annulus
and the  Klein bottle and M\"obius strip projections can be explicitly
computed in many cases when we cannot obtain  detailed information about the
corresponding $n$-point functions in presence of boundaries or crosscaps.
The method is particularly powerful if the completeness  
of the boundary conditions \cite{[98]} is used. Note that modular invariance
of the torus partition function plays also the role of completeness 
condition for the two dimensional fields.   

One starts with a general (not necessary rational) two dimensional theory with isomorphic 
chiral and antichiral observable algebras 
$\cal A$ and $\bar {\cal A}$,
corresponding to a symmetric $X_{ij}=X_{ji}$ torus modular invariant \eref{eq:1.51b}.

To simplify the formulae we shall assume that the theory is rational and 
that the modular invariant is of the 
permutation type \eref{eq:1.56}. This has the advantage 
that one can write all expressions using only chiral labels, while in the general case 
additional degeneracy labels may be needed to distinguish fields with multiplicities
larger than one.

\subsection{Klein bottle projection}

Let us first construct the non-oriented sector. The simplest non orientable
surface, the Klein bottle, can be represented as a cylinder terminating
at two crosscaps. The Klein bottle contribution to the partition function
\index{partition function ! Klein bottle}
 is a linear combination of the  Virasoro characters \cite{[92],[103],[104]}, 
 hence in general it is not a modular invariant.
 In fact there are two distinct expressions for the 
 Klein bottle contribution  called direct and transverse channel
 which are related by the modular $S$ transformation \eref{eq:1.52a}.
 In the string theory language they  correspond to inequivalent choices of time on the worldsheet.
 In the direct channel the Klein bottle contribution is a projection of the torus
 partition function that describes the (anti)symmetry properties
 of the two dimensional fields under the involution 
 $\Omega$ \eref{eq:4.42a}
\begin{equation}
K \ = \  \sum_i \ \chi_i \ K^i \quad  , 
\label{eq:4.54a}
\end{equation}
where the integers 
 $K^i$ satisfy 
\begin{equation}
| K^i | \leq  X_{i i} \qquad   K^i  =  X_{i i} \ ({\rm mod } \ 2) \ .
\label{eq:4.54b}
\end{equation}
Hence for permutation invariants  $K^i$ can take only the values
 $0$ or $ \pm 1$. The  $K_i$ are related (but not necessary equal) to the signs 
 $\epsilon_{i}$ in \eref{eq:4.42a}.

The modular  $S$ transformation  turns \eref{eq:4.54a} 
into the transverse channel, which describes the reflection of the 
two dimensional fields from the two croscaps at the ends of the cylinder.
It has the form 
\begin{equation}
\tilde K \ = \  \sum_i \ \chi_i \ {\Gamma_i}^2 \quad , 
\label{eq:4.55}
\end{equation}
where the reflection coefficients  $\Gamma_i$ are the normalizations
of the one point functions of the two dimensional fields
in front of the crosscap (see equation \eref{eq:4.40}),
so they vanish if $X_{i i^C} = 0$.   

The complete partition function 
in the unoriented case is given by the half sum of the 
torus and direct channel Klein bottle contributions 
\begin{equation}
Z_{\rm unoriented} = {1 \over 2} (Z_T + K) \ .
\label{eq:4.551}
\end{equation}
The multiplicity of a field $\phi_{ij} (= \phi_{ji})$ can be read of 
 the partition function \eref{eq:4.551} as follows 
 (for  a permutation invariant,
 if there are multiplicities the argument applies for each copy of 
 the fields)

- if $ i \neq j$ it is equal to $1/2(X_{ij} +X_{ji})$ and is 
non-negative integer due to the assumption that the torus invariant is symmetric.
Only one combination of the two fields $\phi_{ij}$ and $\phi_{ji}$ remains in the spectrum, 
the other is projected out.

- if $ i = j$ it is equal to $1/2(X_{ii}+K_{i})$ and is 
non-negative integer since  $K_i$ satisfy \eref{eq:4.54b}.
In particular if $X_{ii}=1$ the fields with  $K_i=1$ remain in the 
spectrum, while the ones with $K_i=-1$ are projected out.

If the ground state is degenerate, the Klein bottle projects
out the part antisymmetric under the left-right exchange  rather that  
the whole field.

We shall illustrate the construction on the
example of the non-diagonal $D_5$ 
model of the  $SU(2)$ current algebra with level  $k=6$
with torus partition function \eref{eq:4.53}.
There are two different Klein bottle projections, corresponding to the two choices 
of the signs  $\epsilon_i$ in  \eref{eq:4.42a} \cite{[102]}.
For reasons that will become clear in the next subsection,
we shall distinguish them by the subscripts 
 $r$ and $c$ (for ``real'' and ``complex'') 
\begin{eqnarray}
K_r^{D_5} \ &=& \  \chi_1 + \chi_3 + \chi_5 + \chi_7 - \chi_4 \quad 
\label{eq:4.56a} \\
K_c^{D_5} \ &=& \  \chi_1 + \chi_3 + \chi_5 + \chi_7 + \chi_4 \quad . 
\label{eq:4.56b} 
\end{eqnarray}
Comparison with the values of  $\epsilon_i$ given by 

- $\epsilon_i = 1$ for all  $i$ in the real case 

- $\epsilon_i = (-1)^{i-1}$  in the complex case 

\noindent
shows that there is a relative factor  $(-1)^{2I}$ between  $K_i$ and the signs  $\epsilon_i$ which
comes from the $SU(2)$ structure of the fields.
Indeed the singlet is in the symmetric (antisymmetric) part of the tensor 
product of integer (half integer) isospins. So the real Klein bottle projection
correspond to keeping all singlets, while the complex one
projects out the singlet corresponding to $\chi_4 $.

As an application of these ideas to string theory, let us mention that in \cite{[0Bprime]}
by a non-standard Klein bottle projection has been constructed the 
first tachyon free non-supersymmetric open string model.

\subsection{Annulus partition function}

The spectrum of the boundary fields is described by the annulus
(or cylinder) 
partition function 
\index{partition function ! annulus}
with all possible boundary conditions at the two ends.
Again the partition function is linear in the characters, hence not 
a modular invariant, so there are two distinct expressions for the 
annulus contribution \cite{[93]}. They are called direct and transverse channel
and are related by the modular $S$ transformation \eref{eq:1.52a}.

In the direct channel the annulus partition function counts the 
number of operators that intertwine the boundary conditions at the 
two ends and can be represented as
\begin{equation}
A \ = \  \sum_{i,a,b} \ \chi^i \ A^{a b}_i \  n_a \ n_b \quad ,
\label{eq:4.57}
\end{equation}
where the non-negative integers $A^{a b}_i$ give the 
multiplicities of the boundary fields
$\psi_i^{a b}$. The auxiliary multiplicities $n^a$
associated with the boundaries in open string models 
correspond to the 
introduction of Chan-Paton gauge groups \cite{[105]}, which can be  $U(n)$,
$O(n)$ or $USp(2n)$ \cite{[106]}. In the case of  $U(n)$ 
groups the boundaries can be  oriented, since there are two inequivalent choices
of the fundamental representation, hence the Chan-Paton charges come in 
numerically equal pairs $\bar n = n$. We shall call such charges 
complex. The other two cases, $USp(2n)$ and $O(2n)$,
do not lead to similar identifications and we shall call the 
corresponding charges real. The labels $r$ and $c$ on the
partition functions originate from this interpretation. In applications to 
Statistical Mechanics one may regard \eref{eq:4.57} as a generating function
for the multiplicities of the allowed boundary fields.

The transverse channel, related to \eref{eq:4.57} by a modular $S$ 
transformation, has a very different interpretation. It describes 
the reflection of a two dimensional field from the two boundaries and can 
be represented as  
\begin{equation}
\tilde A = \sum_{i} \ \chi^i \ {\left[ \sum_a \ {\cal B}_{i a} \ n^a \right]}^2
\ .
\label{eq:4.58a}
\end{equation}
Since only fields with charge conjugate 
chiral and antichiral labels can couple to the boundaries
it is again sufficient to specify only the chiral label.
The reflection coefficient ${\cal B}_{i a}$ for the field  
$i$ ($\bar i$) from a boundary  $a$ is proportional to the coefficient
of the identity operator 
in the bulk-to-boundary expansion of the two dimensional field in front 
of the boundary  \eref{eq:4.22a}  
\index{structure constants ! bulk-to-boundary}
\begin{equation}
{\cal B}_{i a} \ = \ { C_{(i,\bar i) 1}^a \alpha^{aa}_1 \over \sqrt{N_{i\bar i}}} \ .
\label{eq:4.58b}
\end{equation}

One can  define  
charge conjugation on the boundary labels. 
It is non-trivial only if the boundaries are oriented (that corresponds 
to complex charges) and is given by the involutive matrix 
$(A_1)_{a b} = (A_1)^{a b}$, such that
\begin{equation}
{A_{i a}}^{b} = \sum_c A_{1 a c} \ {A_{i}}^{c b} \ , 
\label{eq:4.59a}
\end{equation}
hence $(A_1)_a^b = \delta_a^b $.
Let us also assume that the boundaries  are a complete set.
To justify this assumption let us recall that the modular invariance condition 
of the torus partition function plays also the role of completeness 
condition for the two dimensional fields.  
The completeness condition for the boundaries has two equivalent formulations. 
The first one \cite{[98]} is to require that 
the coefficients ${A_{ia}}^b$ satisfy the fusion algebra \index{algebra ! fusion}
\begin{equation}
\sum_b \ {A_{i a}}^{b} \ {A_{j b}}^c \  = \
\sum_k N_{ij}^k \  {A_{k a}}^c   \quad . 
\label{eq:4.60a} 
\end{equation}
Intuitively this relation corresponds to two different ways of counting the
boundary fields.
The second one \cite{[PZ PL]} is to require that 
the boundary states are related to the 
Ishibashi states \eref{eq:4.12} by a unitary transformation,
which in particular implies that they are the same number.

Equation  \eref{eq:4.60a}   
contains only chiral information, so it 
cannot determine completely the multiplicities 
$A_i^{a b}$.  
The two dimensional input is provided by the torus modular invariant 
\eref{eq:1.51a}. In particular if for some $j$ the torus coefficient
 $X_{j j^C}=0$ (where 
$j^C$ is the charge conjugate of $j$) then there is no 
two dimensional field with these labels, so the coefficients  
$C_{(j  j^C) 1}^{a}$ are zero for all $a$. 
Hence, due to 
\eref{eq:4.58b} vanish also all 
 ${\cal B}_{j a}$ and  $\chi_j$ will not contribute to 
\eref{eq:4.58a}. 
After a modular transformation this implies 
\begin{equation}
\sum_i \ A_{i}^{a b} \  S^{i}_j \ = \ 0 \quad  
\label{eq:4.61a}
\end{equation}
for all  $a$ and $b$ and this particular $j$.

Hence we can reformulate the problem of finding the annulus partition function
in the following way: solve over the non-negative integers the two equations 
\eref{eq:4.60a} and \eref{eq:4.61a}.
In general this system may have several solutions, but in  all known cases
fixing also the boundary charge conjugation matrix $(A_1)_{a b}$
determines completely all $A_i^{a b}$, and thus the only freedom is in choosing the
orientation on pairs of boundaries. The proof of this fact in the general case is 
however still a challenging open problem.

As an illustration let us again consider the $D_5$ model 
with torus partition function \eref{eq:4.53}.
In the real charge case  $(A_1)_{a b} = \delta_{a b}$
and the solution is (the labels of the charges correspond to the
first column in \tref{tt2}) 
\begin{eqnarray}
 A_r^{D_5} \ &=&  \  
\chi_1 ( n_1^2 + n_2^2 + n_3^2 + n_4^2 + n_5^2 ) \nonumber \\ 
&+& \
(\chi_2 + \chi_6 )( 2 n_1 n_2  + 2 n_1 n_5 + 2 n_3 n_5 + 2 n_4 n_5) \nonumber \\  
&+&
  \chi_3 ( n_1^2 + 2 n_1 n_3 + 2 n_1 n_4 + 2 n_3 n_4 + 2 n_2 n_5 +
 2 n_5^2)
 \nonumber \\  
&+&  \chi_4 ( 4 n_1 n_5 + 2 n_2 n_3 + 2 n_3 n_5 + 2 n_2 n_4 + 2 n_4 n_5 ) 
\nonumber \\ 
&+&
 \chi_5 (n_1^2  + n_3^2  + n_4^2 + 2 n_5^2 + 2 n_1 n_3 + 2 n_1 n_4 +
 2 n_2 n_5 ) \nonumber \\  
&+&
\chi_7 ( n_1^2 + n_2^2 + n_5^2 + 2 n_3 n_4 )   \quad . 
\label{eq:4.62a} 
\end{eqnarray}
In the complex case the two charges $n_3$ and $n_4$ become a 
complex pair  ${\bar n} = n$  and the solution is
\begin{eqnarray}
 A_c^{D_5} \ &=&  \  
\chi_1 ( n_1^2 + n_2^2 + 2 n {\bar n}+ n_5^2 ) \nonumber \\ 
&+& \ (\chi_2 + \chi_6 )( 2 n_1 n_2  + 2 n_1 n_5 + 2 n n_5 + 2{\bar n} n_5) \nonumber \\ 
&+& \chi_3 ( n_1^2 + n^2 + {\bar n}^2 + 2 n_1 n + 2 n_1 {\bar n}  + 2 n_2
n_5 +
 2 n_5^2)
 \nonumber \\ 
&+& \chi_4 ( 4 n_1 n_5 + 2 n_2 n + 2 n_2 {\bar n} + 2 n n_5 + 2 {\bar n}
n_5 )  \nonumber \\ 
&+& \chi_5 (n_1^2 + 2 n_5^2 + 2 n_1 n + 2 n_1 {\bar n} +
 2 n_2 n_5 + 2 n {\bar n}) \nonumber \\ 
&+& \chi_7 ( n_1^2 + n_2^2 + n_5^2 + n^2 +{\bar n}^2)   \quad .
\label{eq:4.62b} 
\end{eqnarray}
Note that in both cases some boundary fields (corresponding to the $n_5$
charge) have multiplicities equal to two.

\subsection{M\"obius strip projection}

The consistency of the theory in presence of both boundaries and crosscaps
is determined by the M\"obius strip contribution \cite{[92],[103],[104]}.
\index{partition function ! M\"obius strip}
The M\"obius strip can be represented as a cylinder terminating at one 
boundary and at one crosscap. Hence in the transverse channel 
the two dimensional field reflects from the boundary and the crosscap with the same 
reflection coefficients ${\cal B}_{i a}$ and $\Gamma_i$ which enter equations  
\eref{eq:4.55}, \eref{eq:4.58a}
\begin{equation}
\tilde M \ = \  \sum_i \ {\hat \chi}^i \ \Gamma_i \ \left[ \sum_a \ {\cal B}_{i a}
\ n^a
\right]
\quad .
\label{eq:4.63}
\end{equation}
As we have seen there are in general more than one  solutions for both ${\cal B}_{i
a}$ and $\Gamma_i$, so we have to specify also which of these solutions we shall use in 
equation   \eref{eq:4.63}. To determine this we can pass to the direct channel
(by a $P$ transformation,  see equation \eref{eq:4.66} below)
\begin{equation}
M \ = \  \sum_i \ {\hat \chi}^i \ M_i^a \ n_a  \quad ,
\label{eq:4.64a}
\end{equation}
and compare this expression with the annulus partition function
\eref{eq:4.57}. The integer coefficients 
 $M_i^a$ can be interpreted as twists (or projections) 
of the open spectrum and thus have to satisfy
\begin{equation}
M_i^a \ = \ A_i^{a a} \  {\rm ( mod  \  2 )} \  \qquad 
\vert M_i^a \vert  \leq A_i^{a a} \ .
\label{eq:4.64b}
\end{equation}
These equations choose 
consistent pairs of annulus and Klein bottle partition functions.

The natural modular parameter in the direct channel for the M\"obius strip
is  $(\rmi \tau +1)/2$, while in the transverse channel it is  
$(\rmi + \tau) / 2 \tau$. The non vanishing real part of the direct channel modular 
parameter implies that the natural basis of characters  for the M\"obius strip
is 
\begin{equation}
{\hat \chi}_j  \ = \ \rme^{- \rmi \pi (\Delta_j - c/24)} \ 
\chi_j \left({\rmi \tau +1 \over 2} \right) \quad ,
\label{eq:4.65}
\end{equation}
hence the transformation which relates the 
direct and the transverse channel is  given by \cite{[103]}
\begin{equation}
P \ = \ T^{1/2}\ S \ T^2 \ S \ T^{1/2} \quad , 
\label{eq:4.66}
\end{equation}
and satisfies $P^2 = C$. The square root of  $T$ in \eref{eq:4.66} denotes the
diagonal matrix whose  eigenvalues are square roots of the eigenvalues of $T$.

By a formula similar to the Verlinde formula \eref{eq:1.59} one can define 
the coefficients ${Y_{ij}}^k$ \cite{[101]}
\begin{equation}
 {Y_{ij}}^k \ = \ \sum_{\ell} \ { S_{i \ell} \ P_{j \ell} \ P_{k
\ell}^{\dagger} 
\over S_{1 \ell}}  \quad , 
\label{eq:4.67}
\end{equation}
which are  integers \cite{[integer Y],[Frobenius]}
and satisfy the fusion algebra \index{algebra ! fusion}
\begin{eqnarray}
\sum_l {Y_{im}}^l  {Y_{jl}}^n \ & =& \ \sum_{\ell} {N_{i j}}^{\ell}
 {Y_{\ell m}}^n  \quad  \label{eq:4.68a} \\
 \ms
\sum_i Y_{ijk} {Y^i}_{lm} \ & =& \ \sum_i Y_{ijm} {Y^i}_{lk} \quad . 
\label{eq:4.68b} 
\end{eqnarray}

The complete partition function in the unoriented open sector is  
\begin{equation}
Z_{open} = { 1 \over 2 } (A \pm M) \quad . 
\label{eq:4.681}
\end{equation}
Its integrality is guaranteed by the conditions
\eref{eq:4.64b}.
Note that the overall sign of the  
M\"obius strip projection is not determined by the conformal theory. 
In open string models this sign is fixed by the tadpole cancellation conditions \cite{[tadpoles]}
and determines the gauge group.

The completeness condition \eref{eq:4.60a}
implies two relations between the integer
coefficients in  the direct channel 
partition functions $A$, $M$ and $K$  and the $Y$ tensor \eref{eq:4.67}
\begin{eqnarray}
 \sum_b {A_i}^{ab} M_{jb} \ &=& 
 \ \sum_l {Y_{ij}}^l {M_l}^a \quad  
\label{eq:4.70b} \\
\ms
 \sum_b {M_i}^{b} M_{jb} \ &=& 
 \ \sum_l {Y^l}_{ij} K_l \quad  
\label{eq:4.70c} 
\end{eqnarray}
that put very strong constraints 
on $K_i$ and ${M_i}^{a}$ for given ${A_i}^{ab}$
(in all known cases they completely determine them).

Coming back to our example, 
in the non diagonal $D_5$ model there are two 
 consistent choices for  the annulus and Klein bottle partition 
functions, namely  the pairs with the same subscript  ($r$ or $c$). 
The two M\"obius strip projections are correspondingly
\begin{eqnarray}
M_r^{D_5} &=&  \  {\hat \chi}_1 ( n_1 - n_2 + n_3 +
n_4 - n_5 ) \nonumber \\
 &+&
 \ {\hat \chi}_3 ( - n_1 + 2 n_5 ) \nonumber \\
 &+&
 \ {\hat \chi}_5 (n_1 + n_3 + n_4 ) \nonumber \\
 &+& 
\ {\hat \chi}_7 ( n_1 + n_2 + n_5 )  \quad , 
\label{eq:4.69a} 
\end{eqnarray}
and
\begin{eqnarray}
M_c^{D_5} &=& \  {\hat \chi}_1 (- n_1 + n_2 + n_5 ) \nonumber \\
 &+&
 \ 
{\hat \chi}_3 ( n_1 + n + {\bar n}) \nonumber \\
 &+&
\ {\hat \chi}_5 (n_1 + 2 n_5 ) \nonumber \\
 &+& 
\ {\hat \chi}_7 ( n_1 + n_2 + n + {\bar n} + n_5 )  \quad . 
\label{eq:4.69b} 
\end{eqnarray}
It is instructive to verify that these indeed satisfy the polynomial equations 
and to determine the open spectrum of the models. Note that when  the annulus coefficient
 is equal to $2 n_5^2$, there are two possibilities for the M\"obius strip
coefficient. It can be either $2n_5 = n_5 + n_5$ or $0 = n_5 - n_5$. This 
corresponds to two operators with  equal or opposite symmetrization properties.

\subsection{Solutions for the partition functions}

If the torus modular invariant is given by the charge conjugation matrix
$X = C$ then the number of boundary conditions coincides with the 
number of chiral representations, so we can label both by the same 
label. In this case the standard solution for the annulus was found  
in \cite{[93]}, while the expressions
for the Klein bottle and M\"obius strip were found in \cite{[101]}
\begin{eqnarray}
 A_{i j k} &=& N_{i j k} 
\label{eq:4.71a} \\
\ms
 M_{i j} &=&  Y_{j i 1} 
\label{eq:4.71b} \\
\ms
 K_i &=& Y_{i 1 1} \ .
\label{eq:4.71c} 
\end{eqnarray}
Using the properties of $N_{i j k}$ and
$Y_{i j k}$ it is straightforward to verify that these solutions satisfy all the consistency requirements.
Moreover, the standard Klein bottle projection \eref{eq:4.71c} 
is equal to the  Frobenius-Schur indicator \cite{[Frobenius]}  and 
corresponds to 
keeping all the singlets in the spectrum. 
A modular transformation to the transverse channel gives
\begin{eqnarray}
 \tilde A \ &=&  \ \sum_i \left(  
\sum_j { S_{ij} n^j \over \sqrt{S_{1i}}}
\right)^2 \chi_i
\quad  
\label{eq:4.72a} \\
\ms
 \tilde M &=&  
\sum_i \left(  
\sum_j { P_{1i} S_{ij} n^j \over S_{1i}}
\right) \hat \chi_i 
\quad 
\label{eq:4.72b}
\\
\ms
\tilde K \ &=& \ \sum_i \left( {P_{1i} \over
\sqrt{S_{1i}} } \right)^2 \chi_i
\quad . 
\label{eq:4.72c} 
\end{eqnarray}
Before these general formulae were known, in  
\cite{[95]} the standard Klein bottle and M\"obius 
partition functions for the diagonal case of 
the unitary minimal models had been explicitly constructed.

As a simple example of a non standard solution we shall list also the expressions for the 
second possible solution in the diagonal $SU(2)$ current algebra models of level $k$
denoted by $A_{k+1}$. 
The modular matrices  $S$ and $T$  (we label the fields by $j=2I +1$)  are 
\begin{eqnarray}
S_{jl} \  &=& \  \sqrt{{2 \over k+2}}  \ sin \left({\pi j l  \over k+2}\right) \
\label{eq:C.1} \\
\ms
T_{jl} \  &=&  \ \delta_{jl} \  \rme^{\rmi \pi \left( {j^2 \over 2(k+2)} -
{1 \over 4}\right)} \quad .
\label{eq:C.2}
\end{eqnarray}
The charge conjugation matrix is identity  $C=S^2 = (ST)^3 = 1$. 
The modular matrix  $P = T^{1/2} S T^2 S T^{1/2}$ which satisfies
 $P^2 = C = 1$ is 
\begin{equation}
P_{jl} \  =  \ {2 \over \sqrt{k+2}} \  sin \left( {\pi j l \over 2(k+2)}\right)
 \ (E_k  E_{j+l} + O_k O_{j+l}) \ ,
\label{eq:C.3}
\end{equation}
where $E_n$ and $O_n$ are projectors on $n$ even  and $n$ odd correspondingly.

The standard solution in the diagonal model has $k+1$ real charges and is given by 
(\ref{eq:4.71a}..\ref{eq:4.72c}). The explicit expression for the 
 direct channel Klein bottle projection is 
\begin{equation} 
K_r^{\{A_{k+1}\}} \ = \ \sum_{j=1}^{k+1} \  {{Y}^j}_{11}  \chi_j 
\ = \ \sum_{j=1}^{k+1} \  (-1)^{j-1}  \chi_j 
\end{equation}
hence indeed all singlets are kept in the unoriented spectrum. 

The second solution has also $k+1$ charges (most are in complex pairs)
and in the direct channel is given by \cite{[101]}
\begin{eqnarray}
  K_c^{\{A_{k+1}\}} \ &=& \ \sum_{j=1}^{k+1} \  {{Y}^j}_{k+1,k+1}  \chi_j 
\ = \ \sum_{j=1}^{k+1} \ \chi_j
 \quad  
\label{eq:C.9a} \\
\ms
 A_c^{\{A_{k+1}\}} \ &=& \ \sum_{j,l,m=1}^{k+1} {N_{lm}}^j \chi_{k+2-j} 
n^l n^m\quad  
\label{eq:C.9b} \\
\ms
 M_c^{\{A_{k+1}\}} 
\ &=& \ \sum_{j,l=1}^{k+1} {Y_{l,k+1}}^{j} \hat \chi_j n^l 
 \quad . 
\label{eq:C.9c}
\end{eqnarray}
Note that the Klein bottle projects out the singlets for all
the half integer isospin fields, so they cannot couple
to the identity on the boundaries or the crosscap, hence the corresponding
reflection coefficient should vanish.
After a modular transformation we find in the transverse channel 
\begin{eqnarray}
\tilde K_c^{\{A_{k+1}\}} \ &=& \ \sum_i \left( {P_{k+1,i} \over
\sqrt{S_{1i}} } \right)^2 \chi_i  \quad  
\label{eq:C.10a} \\
\ms 
 \tilde A_c^{\{A_{k+1}\}} \ &=&  \ \sum_i (-1)^{i-1} \ \left(  
\sum_j { S_{ij} n^j \over \sqrt{S_{1i}}}
\right)^2 \chi_i \quad  
\label{eq:C.10b} \\
\ms
\tilde M_c^{\{A_{k+1}\}} \ &=& \ 
\sum_i \left(  
\sum_j { P_{k+1,i} S_{ij} n^j \over S_{1i}}
\right) \hat \chi_i  \quad .
\label{eq:C.10c}
\end{eqnarray}
The vanishing of the reflection coefficients of the fields 
with half integer isospin in \eref{eq:C.10b} implies 
the complex charge identifications
$n_{k+2-i} = \bar n_i = n_i$ for all $i$.

In the $D_{odd}$  models there are again two different choices 
for the Klein bottle projection
(which generalize equations (\ref{eq:4.56a},\ref{eq:4.56b}) 
for $D_{5}$). Both lead to $k/2+2$ charges. 
The corresponding annulus and M\"obius strip partition functions   
are rather involved \cite{[98],[102]}.
The solutions for  $D_{even}$, $E_6$ and $E_8$ 
(with charge conjugation modular invariants if considered as models with 
extended symmetry) are given by the general formulae (\ref{eq:4.71a}..\ref{eq:4.72c}).
The solution for the exceptional case  $E_7$ is given in \cite{[102]}.
In the $D_{even}$ and $E$ models one can study also 
boundary conditions that do not respect the extended
symmetry of the bulk model, but only the $SU(2)$ symmetry.
The corresponding solutions are given in \cite{[PZNP]}.

Many other solutions have been found. Let us note only the general formulae in 
\cite{[FOE]} where the partition functions for all 
simple currents  modular invariants  are given.

\section*{Acknowledgments}
It is a pleasure to thank  the organizers of the 4th SIGRAV 
Graduate School and 2001 School on Algebraic 
Geometry and Physics  ``Geometry and Physics of Branes'' at Como for 
giving me the opportunity to present these lectures.
This work was supported in part
by the EEC contracts HPRN-CT-2000-00122
and HPRN-CT-2000-00148 and by the INTAS contract 99-1-590.

\vfill
\eject


\end{document}